\documentclass[lettersize,journal]{IEEEtran}
\usepackage{amsmath,amsfonts}
\usepackage{array}
\usepackage{textcomp}
\usepackage{stfloats}
\usepackage{url}
\usepackage{verbatim}
\usepackage{comment}
\usepackage{cite}
\usepackage{bm}  
\usepackage{float} 
\usepackage{placeins} 
\usepackage{amsmath,amssymb,amsfonts}
\usepackage{graphicx}
\usepackage{textcomp}
\usepackage{xcolor}
\usepackage{caption}
\usepackage{hyperref}
\usepackage{cite}
\usepackage{multirow}
\usepackage{url}
\usepackage{amsmath,amssymb,amsfonts}
\usepackage{graphicx}
\usepackage{textcomp}
\usepackage{caption}
\usepackage{multicol}
\usepackage{xcolor}
\usepackage{subcaption}
\usepackage{array}
\usepackage{float}
\usepackage{lipsum}
\usepackage{algorithm}
\usepackage{algorithmic}
\usepackage{graphicx}
\usepackage{textcomp}
\usepackage{scalerel}
\usepackage{cite}
\usepackage{amsmath,amssymb,amsfonts}
\usepackage{algorithmic}
\usepackage{graphicx}
\usepackage{textcomp}
\usepackage{caption}
\usepackage{multicol}
\usepackage{esdiff}
\usepackage{cleveref}
\usepackage{etoolbox}
\usepackage{cases}
\usepackage{mathtools}
\usepackage{comment}
\usepackage{tikz}
\usetikzlibrary{arrows.meta,positioning,calc}
\usepackage{multirow}
\usepackage{graphicx}
\usepackage{xcolor}

\newtheorem{remark}{Remark}

\usepackage[none]{hyphenat} 
\usepackage{microtype}      
\allowdisplaybreaks
\begin{document}

    \title{Dynamic Quantum-Assisted Co-Design of Controller and Lyapunov Candidate Parameters for Nonlinear Systems}
	
	\author{Milad Hasanzadeh  \textit{Graduate Student Member, IEEE}, Amin Kargarian, \textit{Senior Member, IEEE}, and Mehdi Farasat, \textit{Senior Member, IEEE}
\thanks{This work was supported by the National Science Foundation under Grant ECCS-2312086.
			
			The authors are with the Electrical and Computer Engineering Department, Louisiana State University, Baton Rouge, LA 70803 USA (email: mhasa42@lsu.edu, kargarian@lsu.edu, mfarasat@lsu.edu).}}

\maketitle

\begin{abstract}
This paper proposes a dynamic quantum-assisted co-design framework for nonlinear closed-loop systems in which controller parameters and Lyapunov candidate parameters are redesigned jointly at successive decision epochs.  Unlike conventional nonlinear control designs that typically tune controller gains offline and perform stability analysis separately, the proposed method embeds performance improvement and evaluation of sampled Lyapunov conditions through penalties within a unified online optimization loop. The main novelty is a two-step computational structure that first contracts the continuous admissible search region around the current operating condition using a Black-Hole calibration procedure and then constructs a finite binary representation only over this calibrated region. The encoded objective is obtained from sampled nonlinear closed-loop evaluations and approximated by a local quadratic pseudo-Boolean surrogate, enabling an Ising-type Hamiltonian representation suitable for quantum-assisted optimization. Quantum imaginary time evolution is then used to explore the encoded Hamiltonian, and the resulting candidate bitstrings are decoded into continuous controller and Lyapunov parameters. To reduce dependence on the surrogate model, the decoded candidates are re-evaluated using the original nonlinear closed-loop cost and Lyapunov penalties before the final update is applied. The framework can accommodate sampled forms of different Lyapunov decay specifications by modifying the corresponding penalty and is numerically evaluated on first-order nonlinear consensus, second-order nonlinear consensus, and induction motor drive control examples. The implementation code used to generate the reported results is available at \href{https://github.com/LSU-RAISE-LAB/DQCLS-NS}{GitHub}.
\end{abstract}


\section{Introduction}
\label{sec:introduction}
\IEEEPARstart{N}{onlinear} dynamical systems arise in many engineering applications, including electric drives, robotics, autonomous systems, and large-scale interconnected processes \cite{khalil2002nonlinear,clarke1997asymptotic}. Their strong state dependence, multiple operating regimes, and complex transients make analysis and control challenging. Among available nonlinear-control tools, this work focuses on Lyapunov-based methods because they provide a constructive framework for stability certification and feedback design. Practical implementations must also maintain performance under uncertainty, disturbances, and changing operating conditions \cite{rawlings1994nonlinear,ang2005pid}, motivating methods that jointly address stability and closed-loop performance \cite{khalil2002nonlinear,ang2005pid}. Despite this foundation, nonlinear controllers are often designed offline for nominal conditions, with stability verified separately using analytical Lyapunov tools \cite{khalil2002nonlinear}. Fixed parameters can become conservative under uncertainty and changing operating regimes \cite{mohammadpour2012control}. Gain scheduling and piecewise designs partially address this issue, but require predefined operating regions, controllers, and stability certificates, which may be poorly matched to the current closed-loop behavior \cite{mohammadpour2012control,johansson2003piecewise,hasanzadeh2025finite}.

These limitations have motivated increasing interest in online redesign and dynamic retuning strategies, in which controller updates are performed during operation so that the closed-loop law can respond to the current operating condition rather than rely on a fixed offline design \cite{mayne2000constrained,hasanzadeh2024distributed1}. In this setting, adaptive control updates parameters online in response to uncertainty and changing plant behavior, gain scheduling and parameter varying approaches modify the applied controller according to measurable operating point information, nonlinear model predictive control repeatedly solves finite horizon optimization problems to account for nonlinear dynamics and constraints, and more recent data-driven updates use measured trajectories to improve closed-loop behavior without relying on a nominal model \cite{leith2000survey,mayne2000constrained,hou2013model}. Collectively, these online redesign methods have expanded the practical reach of nonlinear control by enabling parameter adaptation, constraint handling, and improved responsiveness to regime changes, disturbances, and state-dependent nonlinear behavior as the system evolves \cite{mayne2000constrained,camacho2007constrained,hasanzadeh2024distributed3}. However, this flexibility comes at a substantial computational cost because the redesign problem must be solved repeatedly in closed loop, is typically nonlinear and nonconvex, depends on the current state and prediction model, and generally has no closed-form analytical solution. Therefore, each update usually requires iterative numerical optimization, which can be difficult to execute within real-time control intervals \cite{johansen2011introduction,lucas2014ising}. Moreover, in most existing frameworks, the primary emphasis remains on controller redesign itself, whereas the associated stability analysis is carried out offline, embedded through pre-established assumptions, or verified separately from the online update mechanism \cite{mayne2000constrained,rawlings2020model}. This is because jointly redesigning the controller parameters and the Lyapunov candidate parameters online can improve closed-loop performance while adapting the Lyapunov candidate to the current operating condition. However, this makes the underlying problem more computationally demanding. This added burden has limited the adoption of such co-design strategies in practice. At the same time, advances in computational architectures motivate online co-design frameworks that leverage emerging computing tools while improving closed-loop performance through dynamic redesign \cite{nielsen2010quantum,farhi2014quantum,motta2020determining}.

In parallel with these developments, quantum computing has emerged as a promising computational paradigm for hard optimization and combinatorial search problems that are difficult to address with conventional architectures \cite{hasanzadeh2025admm,berberich2024quantum,hasanzadeh2025admm2}. In particular, optimization problems expressed through binary encodings, Ising models, or quadratic unconstrained binary optimization (QUBO) formulations have attracted significant attention because they admit a natural quantum compatible formulation in terms of Hamiltonian energy minimization \cite{hasanzadeh2026dynamic}. {A related development is the quantum nonlinear model predictive control framework of~\cite{novara2024quantum}, which converts a finite horizon control input optimization problem into a QUBO suitable for quantum annealing.} This encoded optimization viewpoint is relevant when a continuous design problem can be approximated locally by a finite binary search space, thereby enabling the use of quantum routines that operate on discrete decision variables \cite{montanaro2016quantum}. Although currently available quantum hardware remains limited in scale, noise tolerance, and real-time deployability, the future potential of quantum optimization continues to motivate computational frameworks that can map challenging online decision problems into forms amenable to emerging quantum tools \cite{preskill2018quantum,bharti2022noisy}. From this perspective, developing quantum compatible formulations for demanding online control redesign problems is not only interesting, but also timely in light of ongoing progress in quantum algorithms, architectures, and optimization design \cite{farhi2014quantum,motta2020determining,bharti2022noisy}.

Despite recent advances, a gap remains in online nonlinear co-design using quantum optimization \cite{mayne2000constrained,rawlings2020model,farhi2014quantum,motta2020determining}. Existing methods generally do not jointly update controller and {Lyapunov candidate parameters} or accommodate multiple Lyapunov decay specifications within one framework \cite{hou2013model,bhat2000finite,polyakov2011nonlinear}. 
This paper addresses this gap through a dynamic co-design framework whose contribution encompasses joint parameter redesign, calibration before encoding, and quantum compatible optimization using QITE \cite{hatamlou2013black,motta2020determining}. The main contributions are:

\begin{itemize}
    \item We jointly update controller and Lyapunov candidate parameters at each decision epoch, integrating performance optimization and evaluation of sampled Lyapunov conditions rather than relying on a fixed offline candidate.

    \item We calibrate the continuous search region using Black-Hole search before binary encoding, retaining local resolution with fewer binary variables.

    \item We construct a local QUBO/Ising surrogate from nonlinear performance and Lyapunov evaluations, search it using QITE, and evaluate decoded candidates with the original objective. This avoids application specific symbolic reformulations and supports different nonlinear plants and Lyapunov specifications, with the same workflow demonstrated on three control problems.
\end{itemize}

The remainder of this paper is organized as follows. Section~\ref{sec:problem_statement} presents the problem formulation. Section~\ref{sec:proposed_work} develops the proposed two-step dynamic quantum co-design framework. Section~\ref{sec:simulation_results} reports simulation results on several nonlinear control examples. Section~\ref{sec:conclusion} concludes the paper.

\section{Problem Statement}
\label{sec:problem_statement}
This section formulates the nonlinear control problem, including the parametric controller and Lyapunov candidate, the dynamic co-design of their parameters, and the incorporation of different stability specifications around the equilibrium.

\subsection{Nonlinear Closed-Loop Systems}
\label{subsec:general_nonlinear_system}

Consider a nonlinear dynamical system of the form
\begin{equation}
    \dot{x}(t)=f\big(x(t),u(t)\big),
    \label{eq:general_open_loop}
\end{equation}
where \(x(t)\in\mathbb{R}^{n}\) is the state and \(u(t)\in\mathbb{R}^{m}\) is the control input. The vector field \(f:\mathbb{R}^{n}\times\mathbb{R}^{m}\to\mathbb{R}^{n}\) is assumed to be sufficiently regular so that solutions exist and are unique over the interval of interest.

We consider a parametric feedback controller of the form
\begin{equation}
    u(t)=\kappa(t)\,{\chi}\big(x(t)\big),
    \label{eq:general_controller}
\end{equation}
{where \(\chi:\mathbb{R}^{n}\to\mathbb{R}^{n_{\chi}}\) is a prescribed vector of nonlinear state-dependent controller basis functions and \(\kappa(t)\in\mathbb{R}^{m\times n_{\chi}}\) is the matrix of tunable controller coefficients to be redesigned online. Thus, \(u(t)\in\mathbb{R}^{m}\) is linear in the coefficients contained in \(\kappa(t)\), but may be nonlinear in the state through \(\chi(x(t))\).} Substituting \eqref{eq:general_controller} into \eqref{eq:general_open_loop} yields the closed-loop dynamics
\begin{equation}
    \dot{x}(t)=F\big(x(t);\kappa(t)\big),
    \label{eq:general_closed_loop}
\end{equation}
where \(F(\cdot)\) is induced by plant dynamics and feedback law.

To analyze the stability of \eqref{eq:general_closed_loop}, we adopt the direct Lyapunov approach and seek a scalar candidate that is positive definite around the equilibrium and whose derivative along the closed-loop trajectories satisfies the required decay condition. In this work, the Lyapunov candidates are parameterized as
\begin{equation}
{
V(x;\theta)=\sum_{j=1}^{n_{\theta}}\theta_j\,\varphi_j(x),
\qquad
\theta\in\Theta\subset\mathbb{R}^{n_{\theta}},
}
\label{eq:general_lyapunov_candidate}
\end{equation}
{where \(n_{\theta}\) denotes the number of Lyapunov basis functions and, equivalently, the dimension of the coefficient vector \(\theta\).} {The candidate \(V\) is linear in the coefficients \(\theta_j\), but is generally nonlinear in the state \(x\) through the basis functions \(\varphi_j(x)\).}

To enable online adaptation, the redesign is performed at discrete decision epochs
\begin{equation}
    0=t_{0}<t_{1}<\cdots<t_{k}<\cdots,
\end{equation}
with \(t_{k+1}-t_{k}=\Delta t\) for a prescribed redesign interval \(\Delta t>0\). At each epoch \(t_k\), the current closed-loop state \(x(t_k)\) is used to compute the updated parameters
\begin{equation}
\label{eq:kappa}
    \kappa_k\in\mathbb{R}^{n_{\kappa}},
    \qquad
    \theta_k\in\Theta\subset\mathbb{R}^{n_{\theta}},
\end{equation}
{where \(\kappa_k\) denotes the vectorized controller coefficient matrix at \(t_k\).} These parameters are then applied over the interval \([t_k,t_{k+1})\). Hence, the closed-loop system evolves according to
\begin{equation}
    \dot{x}(t)=F\big(x(t);\kappa_k\big),
    \qquad
    t\in[t_k,t_{k+1}),
    \label{eq:piecewise_closed_loop}
\end{equation}
{with the corresponding Lyapunov candidate over this interval given by \(V(x;\theta_k)\). Since \(V(x;\theta_k)\) may change across redesign epochs, the present results provide interval-wise, sample-based evidence and do not constitute a rigorous switched or hybrid system stability guarantee.}

\begin{remark}
Without loss of generality, the stability analysis is carried out around the origin after a standard coordinate shift from the equilibrium point of interest.
\end{remark}

\subsection{Dynamic Optimization Formulation}
\label{subsec:optimization_formulation}

At each decision epoch \(t_k\), the objective is to dynamically co-design the controller parameters \(\kappa_k\) in \eqref{eq:piecewise_closed_loop} and the {Lyapunov candidate parameters} \(\theta_k\) in \eqref{eq:general_lyapunov_candidate}. 
More specifically, \(\kappa_k\) is tuned to improve control performance, while \(\theta_k\) is jointly adjusted using penalties on violations of the prescribed Lyapunov inequalities at the sampled states.

To quantify the quality of a candidate redesign, we consider a dynamic performance index of the form
\begin{equation}
\mathcal{J}_k(\kappa_k,\theta_k)
=
\int_{t_k}^{t_{k+1}}
\left(
w_e \, \|\varepsilon(x(\tau))\|^2
+
w_u \, \|u(\tau)\|^2
\right)d\tau,
\label{eq:general_cost_function}
\end{equation}
where \(w_e>0\) and \(w_u>0\) are user-defined weights, \(u(\tau)\) is the control action induced by the selected controller parameters, and {\(\varepsilon(x)\) is the application-dependent closed-loop error signal. For example, \(\varepsilon(x)=Lx\) measures disagreement in consensus, while in tracking problems it represents the state or output tracking error.} The first term measures control performance, while the second penalizes excessive control effort.

Accordingly, the dynamic co-design problem is in the form
\begin{subequations}
\label{eq:general_codesign_problem}
\begin{align}
\min_{\kappa_k,\theta_k}\quad
& \mathcal{J}_k(\kappa_k,\theta_k)
\label{eq:general_codesign_problem_obj}
\\
\text{s.t.}\quad
& \dot{x}(t)=F\big(x(t);\kappa_k\big), \qquad t\in[t_k,t_{k+1})
\label{eq:general_codesign_problem_dyn}
\\
& V\big(0;\theta_k\big)=0
\label{eq:general_codesign_problem_Vzero}
\\
& V\big(x(t);\theta_k\big)>0, \qquad x(t)\neq 0
\label{eq:general_codesign_problem_Vpos}
\\
& \dot{V}\big(x(t);\kappa_k,\theta_k\big)<0, \qquad x(t)\neq 0
\label{eq:general_codesign_problem_Vdotneg}
\\
& c\big(x(t),u(t),\kappa_k,\theta_k\big)\leq 0.
\label{eq:general_codesign_problem_constraints}
\end{align}
\end{subequations}
In \eqref{eq:general_codesign_problem}, constraint \eqref{eq:general_codesign_problem_dyn} enforces the nonlinear closed-loop dynamics under the controller parameters selected at epoch \(t_k\). Constraints \eqref{eq:general_codesign_problem_Vzero}--\eqref{eq:general_codesign_problem_Vpos} impose the standard positive-definiteness requirements on the Lyapunov candidate. Constraint \eqref{eq:general_codesign_problem_Vdotneg} enforces the required decrease condition of the Lyapunov function along the closed-loop trajectories. Constraint \eqref{eq:general_codesign_problem_constraints} represents other application-dependent requirements, such as actuator bounds, state limitations, safety constraints, structural conditions, and other control specifications.

The optimization problem in \eqref{eq:general_codesign_problem} is generally nonconvex and difficult to solve directly for nonlinear systems. Moreover, the Lyapunov conditions must hold over a domain of interest rather than at a single point, which makes the problem inherently infinite dimensional. To obtain a tractable formulation, we adopt a pointwise sampling procedure over the domain induced by the current operating condition. Let \(\mathcal{S}_k=\{x_k^{(1)},x_k^{(2)},\dots,x_k^{(N_s)}\}\) denote the set of sampled points used at epoch \(t_k\). Then, the sample-based co-design problem can be written as
\begin{subequations}
\label{eq:sample_based_codesign_problem}
\begin{align}
\min_{\kappa_k,\theta_k}\quad
& \widehat{\mathcal{J}}_k(\kappa_k,\theta_k)
\label{eq:sample_based_codesign_problem_obj}
\\
\text{s.t.}\quad
& \dot{x}^{(i)}=F\big(x^{(i)};\kappa_k\big), \qquad i=1,\dots,N_s
\label{eq:sample_based_codesign_problem_dyn}
\\
& V\big(0;\theta_k\big)=0
\label{eq:sample_based_codesign_problem_Vzero}
\\
& V\big(x^{(i)};\theta_k\big)>0, \qquad x^{(i)}\neq 0,\;\; i=1,\dots,N_s
\label{eq:sample_based_codesign_problem_Vpos}
\\
& \dot{V}\big(x^{(i)};\kappa_k,\theta_k\big)<0, \qquad x^{(i)}\neq 0,\;\; i=1,\dots,N_s
\label{eq:sample_based_codesign_problem_Vdotneg}
\\
& c\big(x^{(i)},u^{(i)},\kappa_k,\theta_k\big)\leq 0, \qquad i=1,\dots,N_s.
\label{eq:sample_based_codesign_problem_constraints}
\end{align}
\end{subequations}
where \(\widehat{\mathcal{J}}_k(\kappa_k,\theta_k)\) denotes the sampled approximation of the performance index. In this form, the original continuous co-design problem is converted into a finite dimensional sample-based Lyapunov design problem in which the controller parameters \(\kappa_k\) and the candidate parameters \(\theta_k\) are jointly selected using finitely many evaluations of the nonlinear closed-loop system and the associated Lyapunov inequalities.

\subsection{Stability Specifications}
\label{subsec:stability_specifications}

In the implementations considered in this paper, the basis functions are selected such that \(\varphi_j(0)=0\) for all \(j\), which enforces \(V(0;\theta_k)=0\) structurally. If a more general Lyapunov basis is used, this condition can be included through an additional penalty term \(\rho_0\|V(0;\theta_k)\|^2\).

The sampled Lyapunov requirement associated with \eqref{eq:sample_based_codesign_problem_Vpos}--\eqref{eq:sample_based_codesign_problem_Vdotneg} depends on the desired convergence behavior and the application under study. In all cases, \eqref{eq:sample_based_codesign_problem_Vpos} is retained as the positivity requirement on the Lyapunov candidate at the sampled states, whereas the sampled decay condition is adjusted according to the desired stability notion.

In the present framework, Lyapunov conditions are evaluated at sampled states through penalty terms; this does not ensure their satisfaction or establish stability over a continuous domain or across parameter updates. For sampled evaluation of an asymptotic stability decay condition \cite{khalil2002nonlinear}, we consider

\begin{equation}
V(x^{(i)};\theta_k)>0,\qquad x^{(i)}\neq 0,
\label{eq:asymptotic_Vpos}
\end{equation}
\begin{equation}
\dot{V}(x^{(i)};\kappa_k,\theta_k)<0,\qquad x^{(i)}\neq 0,
\label{eq:asymptotic_Vdot}
\end{equation}
for \(i=1,\dots,N_s\). These inequalities specify positivity and strict decrease at the sampled states.

For sampled evaluation of an exponential decay condition with rate parameter \(\alpha\) \cite{khalil2002nonlinear}, we consider
\begin{equation}
\dot{V}(x^{(i)};\kappa_k,\theta_k)+\alpha V(x^{(i)};\theta_k)<0,
\label{eq:exponential_Vdot}
\end{equation}
for \(x^{(i)}\neq 0\), \(i=1,\dots,N_s\), and \(\alpha>0\).
Unlike the asymptotic decay condition, \eqref{eq:exponential_Vdot} specifies a decay rate through the parameter \(\alpha\).

For sampled evaluation of a finite time Lyapunov decay condition \cite{bhat2000finite}, we consider
\begin{equation}
\dot{V}(x^{(i)};\kappa_k,\theta_k)+cV(x^{(i)};\theta_k)^{\gamma}\leq 0,
\label{eq:finite_time_Vdot}
\end{equation}
for \(x^{(i)}\neq 0\), \(i=1,\dots,N_s\), where \(c>0\) and \(0<\gamma<1\). The fractional power in \eqref{eq:finite_time_Vdot} corresponds to finite time convergence, with a settling time bound that depends on the initial condition.

For sampled evaluation of a fixed time Lyapunov decay condition \cite{polyakov2011nonlinear}, we consider
\begin{equation}
\dot{V}(x^{(i)};\kappa_k,\theta_k)
+aV(x^{(i)};\theta_k)^{p}
+bV(x^{(i)};\theta_k)^{q}\leq 0,
\label{eq:fixed_time_Vdot}
\end{equation}
for \(x^{(i)}\neq 0\), \(i=1,\dots,N_s\), where \(a,b>0\), \(0<p<1\), and \(q>1\).
The two powers in \eqref{eq:fixed_time_Vdot} correspond to fixed time convergence, whose settling time bound is independent of the initial condition.

Accordingly, the sample-based co-design problem in \eqref{eq:sample_based_codesign_problem} keeps the positivity requirement on \(V\) and replaces \eqref{eq:sample_based_codesign_problem_Vdotneg} with the decay inequality associated with the desired stability notion. The main computational workflow remains unchanged across these stability specifications, but each case requires a problem-specific Lyapunov candidate, decay parameters, penalty tuning, sampling region, and the assumptions associated with the selected stability notion. Accordingly, the framework can accommodate these sampled decay forms.

\section{Two-Step Dynamic Quantum Co-Design Framework}
\label{sec:proposed_work}
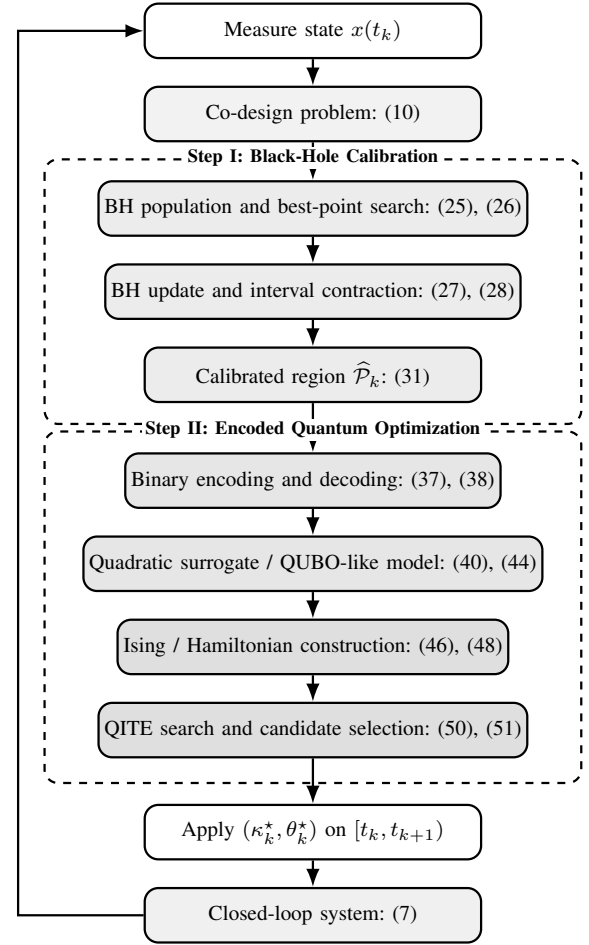
\begin{figure}[!t]
\centering
\begin{tikzpicture}[>=Latex, line width=0.85pt, font=\footnotesize]

\tikzstyle{block}=[draw=black, rounded corners=1.8mm, align=center,
minimum width=4.45cm, minimum height=0.72cm, inner sep=2.2pt]
\tikzstyle{term}=[block, fill=white]
\tikzstyle{blackb}=[block, fill=gray!10]
\tikzstyle{greenb}=[block, fill=gray!15]
\tikzstyle{orangeb}=[block, fill=gray!20]
\tikzstyle{purpleb}=[block, fill=gray!25]
\tikzstyle{grayb}=[block, fill=gray!12]

\node[term]    (state)   at (0.35,0.0)    {Measure state $x(t_k)$};

\node[blackb]   (problem) at (0.35,-1.10)  {Co-design problem: \eqref{eq:sample_based_codesign_problem}};

\node[greenb]  (bhpop)   at (0.35,-2.35)
{BH population and best-point search \cite{hatamlou2013black}:\\
\eqref{eq:bh_population}, \eqref{eq:black_hole_center}};

\node[greenb]  (bhupd)   at (0.35,-3.45)  {BH update and interval contraction: \eqref{eq:bh_update_rule}, \eqref{eq:bh_interval_update}};
\node[greenb]  (region)  at (0.35,-4.55)  {Calibrated region $\widehat{\mathcal P}_k$: \eqref{eq:calibrated_search_region}};

\node[orangeb] (binary)  at (0.35,-5.95)  {Binary encoding and decoding: \eqref{eq:calibrated_full_binary_encoding}, \eqref{eq:calibrated_decoding_operator}};
\node[orangeb] (qubo)    at (0.35,-7.05)  {Quadratic surrogate / QUBO-like model: \eqref{eq:calibrated_quadratic_surrogate}, \eqref{eq:calibrated_QUBO_like_problem}};
\node[purpleb] (ising)   at (0.35,-8.15)
{Ising / Hamiltonian construction \cite{lucas2014ising}:\\
\eqref{eq:Ising_energy}, \eqref{eq:qite_hamiltonian}};
\node[purpleb] (qite)    at (0.35,-9.25)
{QITE search and candidate selection \cite{motta2020determining}:\\
\eqref{eq:qite_parameter_update}, \eqref{eq:final_qite_selection}};

\node[term]    (apply)   at (0.35,-10.60) {Apply $(\kappa_k^\star,\theta_k^\star)$ on $[t_k,t_{k+1})$};
\node[grayb]   (plant)   at (0.35,-11.70) {Closed-loop system: \eqref{eq:piecewise_closed_loop}};

\draw[->, black] (state.south) -- ++(0,-0.12) -- (problem.north);
\draw[->, black] (problem.south) -- ++(0,-0.12) -- (bhpop.north);
\draw[->, black] (bhpop.south) -- ++(0,-0.12) -- (bhupd.north);
\draw[->, black] (bhupd.south) -- ++(0,-0.12) -- (region.north);
\draw[->, black] (region.south) -- ++(0,-0.12) -- (binary.north);
\draw[->, black] (binary.south) -- ++(0,-0.12) -- (qubo.north);
\draw[->, black] (qubo.south) -- ++(0,-0.12) -- (ising.north);
\draw[->, black] (ising.south) -- ++(0,-0.12) -- (qite.north);
\draw[->, black] (qite.south) -- ++(0,-0.12) -- (apply.north);
\draw[->, black] (apply.south) -- ++(0,-0.12) -- (plant.north);

\draw[->, black] (plant.west) -- (-3.55,-11.70) -- (-3.55,0.0) -- (state.west);

\draw[dashed, rounded corners=1.8mm, black] (-3.20,-1.70) rectangle (3.90,-5.20);
\draw[dashed, rounded corners=1.8mm, black] (-3.20,-5.30) rectangle (3.90,-9.95);

\node[fill=white, inner sep=1pt] at (0.35,-1.68) {\scriptsize\bfseries Step I: Black-Hole Calibration};
\node[fill=white, inner sep=1pt] at (0.35,-5.28) {\scriptsize\bfseries Step II: Encoded Quantum Optimization};

\end{tikzpicture}
\caption{\footnotesize Overview of the proposed online co-design framework}
\label{fig:overview_framework}
\end{figure}

This section presents the proposed two-step dynamic co-design framework for solving \eqref{eq:sample_based_codesign_problem}. The sampled constrained problem is first converted into an unconstrained binary form, followed by the two solution steps, dynamic algorithm, and implementation details. Fig.~\ref{fig:overview_framework} summarizes the overall workflow.

\subsection{QUBO-Like Problem Formulation}
\label{subsec:qubo_formulation}

Recall that, at decision epoch \(t_k\), the sample-based co-design problem is given by \eqref{eq:sample_based_codesign_problem}. For notational convenience, we define the joint design vector as
\begin{equation}
p_k :=
\begin{bmatrix}
\kappa_k \\
\theta_k
\end{bmatrix}
\in \mathbb{R}^{n_p},
\qquad
n_p = n_{\kappa}+n_{\theta}.
\label{eq:joint_design_vector}
\end{equation}

To obtain an optimization form suitable for binary encoding, the constrained sample-based problem in \eqref{eq:sample_based_codesign_problem} is rewritten as a penalized unconstrained problem. In this reformulation, the state trajectory is not treated as an independent optimization variable. Instead, for each candidate design vector \(p_k=(\kappa_k,\theta_k)\), the closed-loop model is forward simulated over the redesign horizon using the corresponding controller parameters \(\kappa_k\). Therefore, the dynamics are satisfied by construction during each objective evaluation, while the Lyapunov and application dependent inequalities are added to the objective through penalty terms:
\begin{equation}
\dot{x}(t)=F\big(x(t);\kappa_k\big), \qquad t\in[t_k,t_{k+1}).
\label{eq:simulated_closed_loop_dynamics}
\end{equation}
The simulated trajectory generated by \eqref{eq:simulated_closed_loop_dynamics} is then used to compute both the performance term and the Lyapunov-based penalty terms in the penalized objective. The remaining constraints are incorporated into the objective through penalty terms. Accordingly, we define the penalized cost
\begin{equation}
\widetilde{\mathcal{J}}_k(p_k)
=
\widehat{\mathcal{J}}_k(p_k)
+
\rho_V \Pi_V(p_k)
+
\rho_{\dot V}\Pi_{\dot V}(p_k)
+
\rho_c \Pi_c(p_k),
\label{eq:penalized_cost}
\end{equation}
where \(\rho_V>0\), \(\rho_{\dot V}>0\), and \(\rho_c>0\) are penalty coefficients.

Using the sampled set \(\mathcal{S}_k=\{x_k^{(1)},x_k^{(2)},\dots,x_k^{(N_s)}\}\), the positivity penalty can be written as
\begin{equation}
\Pi_V(p_k)
=
\sum_{i=1}^{N_s}
\left[
\max\Big(0,\epsilon_V - V(x_k^{(i)};\theta_k)\Big)
\right]^2,
\label{eq:V_penalty}
\end{equation}
where \(\epsilon_V>0\) is a small margin. Likewise, the penalty associated with the Lyapunov decay requirement is defined as
\begin{equation}
\Pi_{\dot V}(p_k)
=
\sum_{i=1}^{N_s}
\left[
\max\Big(0,\Psi(x_k^{(i)};\kappa_k,\theta_k)\Big)
\right]^2,
\label{eq:Vdot_penalty}
\end{equation}
where \(\Psi(\cdot)\) denotes the decay condition for the desired stability specification. For asymptotic stability,
\begin{equation}
\Psi(x_k^{(i)};\kappa_k,\theta_k)
=
\dot{V}(x_k^{(i)};\kappa_k,\theta_k).
\label{eq:psi_asymptotic_repeated}
\end{equation}
Finally, application-dependent constraint penalty is defined as
\begin{equation}
\Pi_c(p_k)
=
\sum_{i=1}^{N_s}
\left\|
\max\Big(0,c(x_k^{(i)},u_k^{(i)},\kappa_k,\theta_k)\Big)
\right\|_2^2.
\label{eq:c_penalty}
\end{equation}

With these definitions, the constrained problem in \eqref{eq:sample_based_codesign_problem} is approximated by the unconstrained optimization problem
\begin{equation}
\min_{p_k\in\mathcal{P}_k}
\;
\widetilde{\mathcal{J}}_k(p_k),
\label{eq:unconstrained_codesign_problem}
\end{equation}
where \(\mathcal{P}_k\subset\mathbb{R}^{n_p}\) denotes the admissible search region for the joint design vector at epoch \(t_k\). Although \eqref{eq:unconstrained_codesign_problem} removes the explicit constraints, it remains nonlinear and nonconvex because the cost is induced by nonlinear closed-loop simulation together with Lyapunov-based penalty terms.

\subsection{Two-Step Dynamic Co-Design Procedure}
\label{subsec:two_step_framework}

The reformulation in \eqref{eq:unconstrained_codesign_problem} is continuous, nonconvex, and generally difficult to solve online. In particular, the controller parameters \(\kappa_k\) and the {Lyapunov candidate parameters} \(\theta_k\) may each have wide admissible ranges. A finite binary encoding of \eqref{eq:unconstrained_codesign_problem} is convenient for constructing the discrete surrogate used in the proposed framework and for obtaining an Ising-type representation compatible with quantum optimization. However, a direct binary encoding over these full ranges of all parameters is inefficient, since a wide interval represented by only a few bits gives a coarse discretization, while a finer discretization requires more bits and rapidly enlarges the encoded problem. For this reason, the proposed framework first calibrates the continuous search region around the part of the design space that is most relevant to the current operating condition, and then constructs the encoded discrete model only over this reduced region. In this way, the number of bits per parameter can be kept small while preserving enough local resolution to represent useful candidate solutions.

\subsubsection{Step I: Black-Hole Calibration}
\label{subsubsec:bh_calibration}

{As shown in Step I of Fig.~\ref{fig:overview_framework}, we employ the established gradient-free Black-Hole metaheuristic \cite{hatamlou2013black} to contract the continuous search region around promising candidates before binary encoding.} Let the current admissible region of the joint design vector be
\begin{equation}
\mathcal{P}_k
=
\left\{
p_k\in\mathbb{R}^{n_p}
\;\middle|\;
\underline p_{k,j}\le p_{k,j}\le \overline p_{k,j},
\;\; j=1,\dots,n_p
\right\}.
\label{eq:current_search_region}
\end{equation}
The goal of the first step is to replace this broad region by a smaller one that is better matched to the current state \(x(t_k)\).

To do this, a population of candidate parameter vectors is generated inside \(\mathcal{P}_k\),
\begin{equation}
\mathcal{P}_k^{\mathrm{pop}}
=
\left\{
p_k^{[1]},\dots,p_k^{[N_{\mathrm{BH}}]}
\right\},
\qquad
p_k^{[r]}\in\mathcal{P}_k,
\label{eq:bh_population}
\end{equation}
where \(N_{\mathrm{BH}}\) denotes the selected population size. The choice of \(N_{\mathrm{BH}}\) follows the usual tradeoff in population-based metaheuristics: a larger population provides better coverage of the admissible region and reduces the likelihood of premature contraction, whereas a smaller population reduces the number of closed-loop simulations required at each redesign epoch. Since the Black-Hole algorithm does not prescribe a universal population size, \(N_{\mathrm{BH}}\) is selected empirically based on the dimension of the joint design vector, the computational budget, and the required redesign rate. Each candidate is evaluated using the penalized objective \(\widetilde{\mathcal{J}}_k(\cdot)\), and the best candidate is selected as
\begin{equation}
p_k^{\star}
=
\arg\min_{p_k^{[r]}\in\mathcal{P}_k^{\mathrm{pop}}}
\widetilde{\mathcal{J}}_k(p_k^{[r]}).
\label{eq:black_hole_center}
\end{equation}
This best point acts as the center of attraction in the Black-Hole procedure.

Let \(p_k^{[r,\nu]}\) denote the \(r\)-th candidate at iteration \(\nu\), and let \(p_k^{\star,\nu}\) denote the best candidate at the same iteration. Then the remaining candidates are updated according to
\begin{equation}
p_{k,j}^{[r,\nu+1]}
=
p_{k,j}^{[r,\nu]}
+
\xi_{k,j}^{[r,\nu]}
\left(
p_{k,j}^{\star,\nu}
-
p_{k,j}^{[r,\nu]}
\right),
\label{eq:bh_update_rule}
\end{equation}
where \(\xi_{k,j}^{[r,\nu]}\in[0,1]\) is a random coefficient. {This attraction step is the core of the Black-Hole search, as it progressively concentrates the population around the current best candidate.} After the update, each entry is projected back to its admissible interval so that feasibility is preserved.

Once the population is updated, the interval of each parameter is recalibrated from the minimum and maximum values observed in the current population,
\begin{equation}
\underline p_{k,j}^{(\nu+1)}
=
\min_{1\le r\le N_{\mathrm{BH}}} p_{k,j}^{[r,\nu+1]},
\qquad
\overline p_{k,j}^{(\nu+1)}
=
\max_{1\le r\le N_{\mathrm{BH}}} p_{k,j}^{[r,\nu+1]}.
\label{eq:bh_interval_update}
\end{equation}
In this way, the interval width
\begin{equation}
w_{k,j}^{(\nu+1)}
=
\overline p_{k,j}^{(\nu+1)}
-
\underline p_{k,j}^{(\nu+1)}
\label{eq:interval_width}
\end{equation}
shrinks gradually around the most promising region found by the current population. If this width becomes smaller than a prescribed threshold \(\delta_j>0\), namely
\begin{equation}
w_{k,j}^{(\nu+1)}\le \delta_j,
\label{eq:freeze_rule}
\end{equation}
then that parameter is considered sufficiently calibrated and need not be contracted further. After a fixed number of iterations, or after all active parameters satisfy \eqref{eq:freeze_rule}, the procedure returns the calibrated region
\begin{equation}
\widehat{\mathcal{P}}_k
=
\left\{
p_k\in\mathbb{R}^{n_p}
\;\middle|\;
\underline{\widehat p}_{k,j}\le p_{k,j}\le \overline{\widehat p}_{k,j},
\;\; j=1,\dots,n_p
\right\}.
\label{eq:calibrated_search_region}
\end{equation}

This completes the first dashed block in Fig.~\ref{fig:overview_framework}. This step does not solve the encoded problem directly; rather, it reduces the continuous parameter domain before discretization so that the subsequent binary encoding is performed over a locally relevant interval.

\subsubsection{Step II: Finite Binary Representation and Ising Construction}
\label{subsubsec:encoded_ising_construction}

The second dashed block in Fig.~\ref{fig:overview_framework} begins by converting the continuous problem over the calibrated region into a finite discrete one. This is done by discretizing each entry of the joint design vector only over its calibrated interval. More specifically, for each design variable \(p_{k,j}\), \(j=1,\dots,n_p\), consider the calibrated interval
\begin{equation}
p_{k,j}\in
\left[
\underline{\widehat p}_{k,j},
\overline{\widehat p}_{k,j}
\right].
\label{eq:calibrated_parameter_interval}
\end{equation}
The \(j\)-th parameter is represented by \(n_{k,j}\) binary variables,
\begin{equation}
b_{k,j,\ell}\in\{0,1\},
\qquad
\ell=1,\dots,n_{k,j},
\label{eq:calibrated_binary_variables}
\end{equation}
and these bits are collected in the substring
\begin{equation}
b_{k,j}
=
\begin{bmatrix}
b_{k,j,1} & b_{k,j,2} & \cdots & b_{k,j,n_{k,j}}
\end{bmatrix}^{\top}
\in\{0,1\}^{n_{k,j}}.
\label{eq:calibrated_binary_substring}
\end{equation}
This binary substring is interpreted as the integer
\begin{equation}
\nu_{k,j}(b_{k,j})
=
\sum_{\ell=1}^{n_{k,j}}
2^{\,n_{k,j}-\ell} b_{k,j,\ell},
\label{eq:calibrated_integer_value}
\end{equation}
and then mapped linearly to the calibrated interval through
\begin{equation}
p_{k,j}(b_{k,j})
=
\underline{\widehat p}_{k,j}
+
\frac{\overline{\widehat p}_{k,j}-\underline{\widehat p}_{k,j}}{2^{n_{k,j}}-1}
\,\nu_{k,j}(b_{k,j}).
\label{eq:calibrated_affine_decoding}
\end{equation}
Hence, each substring \(b_{k,j}\) selects a discrete value of the corresponding design parameter within the reduced interval.

By stacking all substrings, the full binary representation is obtained as
\begin{equation}
b_k
=
\begin{bmatrix}
b_{k,1}^{\top} &
b_{k,2}^{\top} &
\cdots &
b_{k,n_p}^{\top}
\end{bmatrix}^{\top}
\in\{0,1\}^{n_q},
\quad
n_q=\sum_{j=1}^{n_p} n_{k,j},
\label{eq:calibrated_full_binary_encoding}
\end{equation}
where \(n_q\) is the total number of binary variables. The associated decoding operator is denoted by
\begin{equation}
p_k=\widehat{\mathcal{D}}_k(b_k).
\label{eq:calibrated_decoding_operator}
\end{equation}
Therefore, after calibration and discretization, the problem over the continuous region \(\widehat{\mathcal{P}}_k\) is replaced by a finite problem over binary vectors \(b_k\).

Using this encoding, the exact objective evaluated on the binary domain can be written as
\begin{equation}
f_k(b_k)
:=
\widetilde{\mathcal{J}}_k\big(\widehat{\mathcal{D}}_k(b_k)\big).
\label{eq:calibrated_exact_binary_objective}
\end{equation}
This function assigns to each bitstring the penalized closed-loop cost obtained after decoding that bitstring into continuous controller and {Lyapunov candidate parameters} and then simulating the corresponding closed-loop system. In general, \(f_k(b_k)\) is not an exact quadratic function of the binary variables. This arises from nonlinear state evolution, trajectory-dependent cost accumulation, and penalty evaluation. 

To obtain a finite encoded model with a structure suitable for discrete optimization, \(f_k(b_k)\) is approximated by a quadratic pseudo-Boolean surrogate. Specifically, consider the model
\begin{equation}
Q_k(b_k)
=
\beta_{0,k}
+
\sum_{r=1}^{n_q}\beta_{r,k} b_{k,r}
+
\sum_{r=1}^{n_q-1}\sum_{s=r+1}^{n_q}\beta_{rs,k} b_{k,r}b_{k,s},
\label{eq:calibrated_quadratic_surrogate}
\end{equation}
where \(\beta_{0,k}\), \(\beta_{r,k}\), and \(\beta_{rs,k}\) are coefficients to be identified. The constant term \(\beta_{0,k}\) represents the baseline level of the surrogate, the linear terms account for the individual effect of each binary variable, and the quadratic terms account for pairwise interactions between binary variables. In this way, \(Q_k(b_k)\) captures the dominant first-order and pairwise dependence of the objective over the current calibrated binary search region.

The surrogate in \eqref{eq:calibrated_quadratic_surrogate} is identified from sampled evaluations of the exact objective. Let
\begin{equation}
\mathcal{B}_k^{\mathrm{train}}
=
\left\{
b_k^{(1)}, b_k^{(2)}, \dots, b_k^{(N_{\mathrm{tr}})}
\right\}
\subseteq \{0,1\}^{n_q}
\label{eq:calibrated_training_bitstrings}
\end{equation}
denote a set of sampled bitstrings taken from the calibrated binary domain. For each \(b_k^{(j)}\in\mathcal{B}_k^{\mathrm{train}}\), the corresponding exact objective value is computed as
\begin{equation}
y_k^{(j)}
=
f_k\big(b_k^{(j)}\big)
=
\widetilde{\mathcal{J}}_k\big(\widehat{\mathcal{D}}_k(b_k^{(j)})\big).
\label{eq:calibrated_training_targets}
\end{equation}
The coefficients of the quadratic surrogate are then chosen so that \(Q_k(b_k)\) matches these sampled objective values as closely as possible. This can be done by solving the fitting problem
\begin{equation}
\min_{\beta_{0,k},\,\beta_{r,k},\,\beta_{rs,k}}
\;
\sum_{j=1}^{N_{\mathrm{tr}}}
\left(
Q_k\big(b_k^{(j)}\big)-y_k^{(j)}
\right)^2.
\label{eq:calibrated_least_squares_fit}
\end{equation}
Thus, the pseudo-Boolean model is not derived from an exact symbolic reduction of the nonlinear co-design problem. Instead, it is obtained as a data-driven local approximation of the true encoded objective over the calibrated search region.

After this fitting step, the exact discrete problem is replaced by the local QUBO-like problem
\begin{equation}
\min_{b_k\in\{0,1\}^{n_q}}
\;
Q_k(b_k).
\label{eq:calibrated_QUBO_like_problem}
\end{equation}

To convert this binary quadratic model into Ising form, each binary variable is mapped to a spin variable according to
\begin{equation}
b_{k,r}=\frac{1-z_{k,r}}{2},
\qquad
z_{k,r}\in\{-1,+1\},
\qquad
r=1,\dots,n_q.
\label{eq:calibrated_binary_to_spin_map}
\end{equation}
Substituting \eqref{eq:calibrated_binary_to_spin_map} into \eqref{eq:calibrated_quadratic_surrogate} gives
\begin{equation}
E_k(z_k)
=
\eta_{0,k}
+
\sum_{r=1}^{n_q}\eta_{r,k} z_{k,r}
+
\sum_{r=1}^{n_q-1}\sum_{s=r+1}^{n_q}\eta_{rs,k} z_{k,r}z_{k,s},
\label{eq:Ising_energy}
\end{equation}
where \(z_k=[z_{k,1},\dots,z_{k,n_q}]^{\top}\), and the coefficients \(\eta_{0,k}\), \(\eta_{r,k}\), and \(\eta_{rs,k}\) are obtained from the coefficients of the pseudo-Boolean surrogate through the standard binary-to-spin transformation. Equation \eqref{eq:Ising_energy} is the final encoded discrete optimization model used in the proposed framework.

\subsection{Quantum Preliminaries and QITE Co-Design}
\label{subsec:qite_step}

\begin{figure}[!t]
    \centering
    \includegraphics[width=\linewidth]{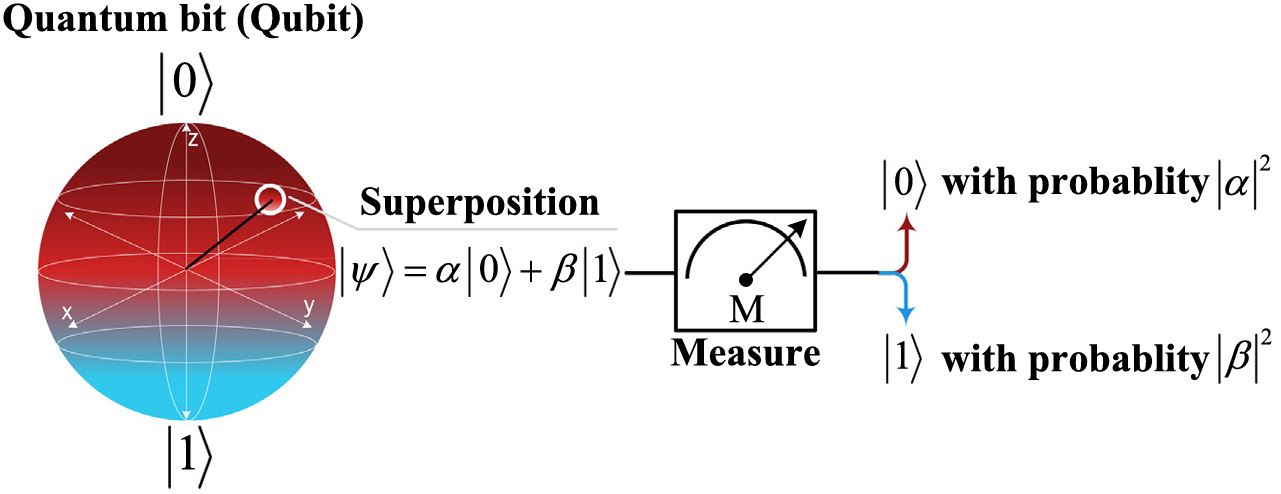}
    \caption{\footnotesize A qubit in superposition and measurement outcomes}
    \label{fig:qubit_measurement}
\end{figure}

Quantum Computing \cite{nielsen2010quantum} operates on qubits, which can exist in a superposition of pure states $|0\rangle=[1,0]^T$ and $|1\rangle=[0,1]^T$. A general single-qubit state is written as $|\psi\rangle = \alpha|0\rangle + \beta|1\rangle$, where $\alpha, \beta \in \mathbb{C}$ and $|\alpha|^2 + |\beta|^2 = 1$. Multiple qubits can be entangled, creating non-classical correlations across subsystems. Quantum operators are unitary matrices denoted by $U$, satisfying $U^\dagger U = I$, where $U^\dagger$ is the Hermitian conjugate of $U$. Applying an operator to a quantum state evolves it as $|\psi_1\rangle = U|\psi_0\rangle$. Typical quantum operators include single-qubit rotations (e.g., $R_x$, $R_y$, $R_z$, Pauli operators) and two-qubit entangling operators such as CNOT. A quantum circuit is a sequence of such operators acting on one or more qubits. A final measurement collapses each qubit into a classical bit (0 or 1), with probabilities determined by the squared magnitudes of the amplitudes. Fig.~\ref{fig:qubit_measurement} illustrates the concept of a qubit in superposition and the probabilistic nature of quantum measurement. It also shows the Bloch sphere representation of a qubit.
Quantum circuits used for optimization can be broadly grouped into fixed and variational circuits. Variational circuits contain parameterized operators whose angles are tunable, whereas fixed circuits do not. QITE is implemented through a variational quantum circuit whose angle parameters are updated during the imaginary-time evolution. {Unlike a conventional variational optimization loop that directly minimizes the Hamiltonian expectation through an external optimizer, QITE determines the parameter evolution from a variational approximation of imaginary-time dynamics.}

The connection between these quantum-computing notions and the proposed co-design is established through the encoded Ising model derived in Section~\ref{subsubsec:encoded_ising_construction}. Each binary variable in the calibrated representation is associated with one qubit, and each candidate bitstring corresponds to a computational-basis state of the quantum register. Therefore, minimizing the encoded surrogate is equivalent to finding a low-energy computational-basis state of a Hamiltonian whose energy values reproduce the Ising objective. This provides the bridge from the QUBO-like surrogate obtained from nonlinear closed-loop evaluations to the QITE-based quantum search used in the proposed framework.

Accordingly, after obtaining the encoded Ising model in \eqref{eq:Ising_energy}, the corresponding quantum Hamiltonian is constructed by replacing each spin variable with the Pauli-\(Z\) operator. For a single qubit, the Pauli-\(Z\) operator is
\begin{equation}
Z=
\begin{bmatrix}
1 & 0\\
0 & -1
\end{bmatrix},
\end{equation}
whose eigenvalues are \(+1\) and \(-1\). Therefore, the encoded Ising energy in \eqref{eq:Ising_energy} can be represented by the Hamiltonian
\begin{equation}
H_k
=
\eta_{0,k} I
+
\sum_{r=1}^{n_q}\eta_{r,k} Z_r
+
\sum_{r=1}^{n_q-1}\sum_{s=r+1}^{n_q}\eta_{rs,k} Z_r Z_s,
\label{eq:qite_hamiltonian}
\end{equation}
where \(I\) is the identity operator, \(Z_r\) acts on qubit \(r\), and \(Z_r Z_s\) denotes the pairwise interaction between qubits \(r\) and \(s\). Since the Hamiltonian is diagonal in the computational basis, each bitstring \(b_k\in\{0,1\}^{n_q}\) corresponds to a basis state, and its associated energy is equal to the value of the encoded Ising objective. Thus, lower-energy basis states correspond to better QUBO solutions, with the ground state giving the global minimum over the encoded search space.

Accordingly, the encoded co-design problem is reduced to the search for a low-energy, ideally ground-state, solution of \(H_k\). In this work, this task is carried out using QITE \cite{motta2020determining}. QITE is motivated by imaginary time evolution, under which the high-energy components of a quantum state are attenuated relative to the low-energy components. As a result, the evolved state is driven toward the ground-state subspace of the Hamiltonian. {Conceptually, the normalized imaginary time state is proportional to \(e^{-\tau H_k}|\psi(0)\rangle\), where increasing the imaginary time \(\tau\) progressively suppresses eigenstate components associated with larger eigenvalues. Because this evolution is nonunitary, QITE approximates it through a sequence of unitary updates within a chosen variational ansatz.}

In the present implementation, the quantum state is restricted to a parameterized ansatz family of the form
\begin{equation}
|\psi(\vartheta)\rangle = U(\vartheta)\,|0\rangle,
\label{eq:qite_ansatz_state}
\end{equation}
where \(U(\vartheta)\) is a fixed parameterized circuit and \(\vartheta\) is the vector of circuit parameters. The ansatz is selected as a hardware efficient circuit acting on the qubits associated with the encoded design variables. Its circuit topology remains fixed, while the parameters are updated throughout the QITE iterations. {The parameterized single-qubit rotations provide state flexibility, while the entangling gates allow correlations among the encoded binary variables to be represented.}

QITE computes the evolution in the space of ansatz parameters so that the state \( |\psi(\vartheta)\rangle \) approximates the imaginary-time evolution generated by \(H_k\). At each iteration, the current ansatz state is prepared, the required expectation values associated with \(H_k\) are evaluated, and a local linear system is formed to determine the parameter update. {This linear system follows from the variational projection of the imaginary-time derivative onto the tangent space of the ansatz and determines the parameter direction that best approximates the desired nonunitary evolution within the selected circuit family.} Denoting this update by \(\Delta\vartheta\), the parameters are modified according to
\begin{equation}
\vartheta \leftarrow \vartheta + \Delta\vartheta.
\label{eq:qite_parameter_update}
\end{equation}
This procedure is repeated for a prescribed number of imaginary-time steps, yielding a final parameter vector \(\vartheta^\star\) and the corresponding state \( |\psi(\vartheta^\star)\rangle \), which approximates a low-energy solution of the encoded Hamiltonian. {The resulting solution quality depends on the ansatz expressiveness, the imaginary time step size, and the number of evolution steps.} This QITE-based search corresponds to the final block of Step II in Fig.~\ref{fig:overview_framework}. The role of QITE is not to replace the original nonlinear closed-loop evaluation, but to explore the encoded Hamiltonian and generate promising candidate bitstrings for subsequent exact re-evaluation.

\begin{figure}[!t]
    \centering
    \includegraphics[width=\columnwidth]{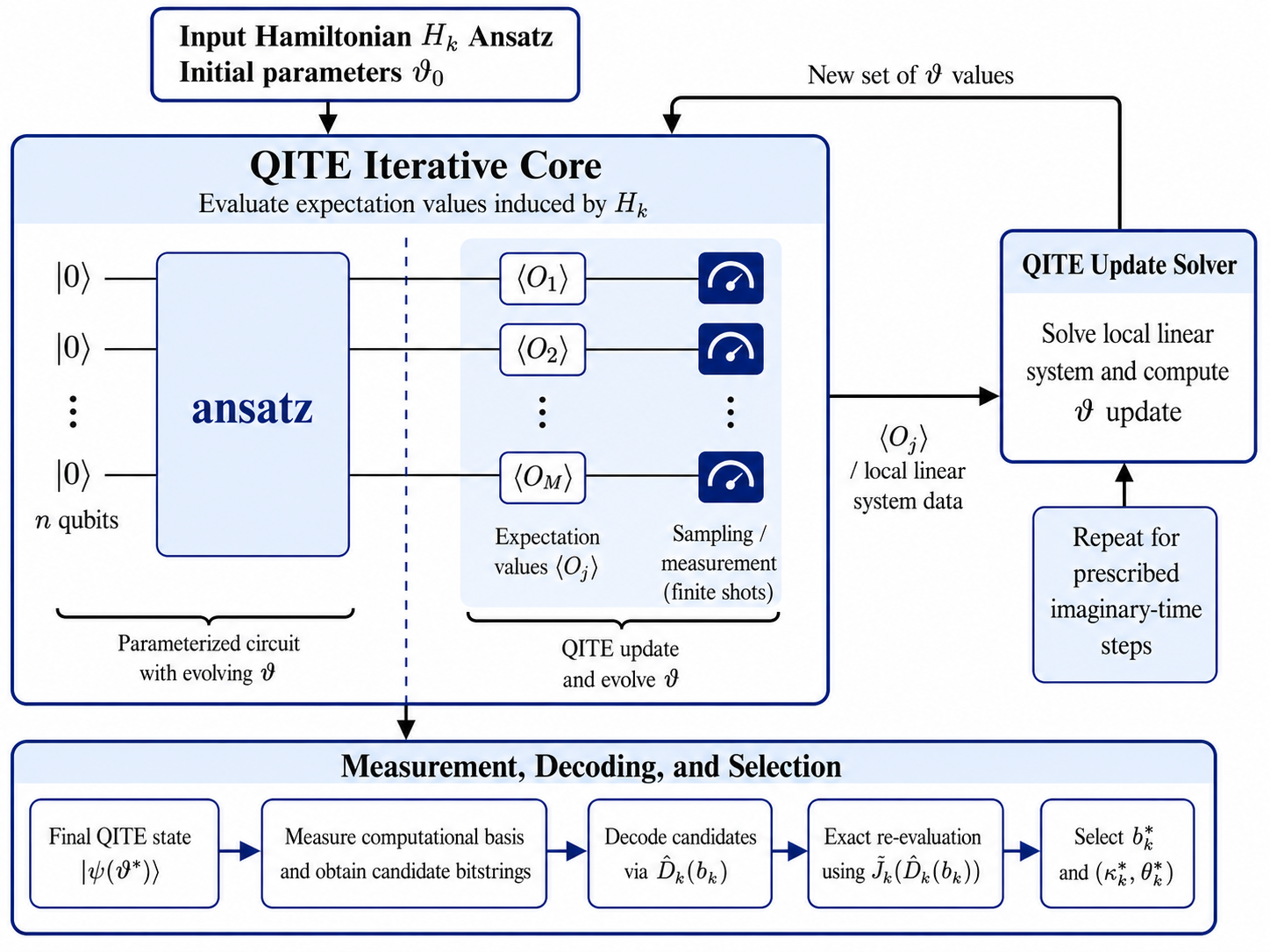}
    \caption{Overview of the QITE-based encoded optimization.}
    \label{fig:qite_workflow}
\end{figure}

The final quantum state is then measured in the computational basis to generate candidate bitstrings. Let \(\mathcal{B}_k^{\mathrm{cand}}\subseteq \{0,1\}^{n_q}\) denote the set of candidate bitstrings selected from the most probable measurement outcomes. {Repeated measurements provide an empirical probability distribution over the encoded solutions, from which the highest-probability distinct bitstrings are retained as candidates.} Each candidate bitstring is decoded into the corresponding continuous design vector through \eqref{eq:calibrated_decoding_operator}, and its exact penalized cost is evaluated using \eqref{eq:penalized_cost}. The final encoded solution is selected as
\begin{equation}
b_k^\star
=
\arg\min_{b_k\in\mathcal{B}_k^{\mathrm{cand}}}
\widetilde{\mathcal{J}}_k\big(\widehat{\mathcal{D}}_k(b_k)\big).
\label{eq:final_qite_selection}
\end{equation}
The selected bitstring \(b_k^\star\) is then decoded to obtain the controller and {Lyapunov candidate parameters} applied at the current redesign epoch. In this way, QITE provides the quantum optimization step of the proposed two-step dynamic co-design framework. {The final exact re-evaluation prevents the selection from relying solely on the fitted surrogate energy and ensures that the applied parameters are compared using the original nonlinear performance and Lyapunov criteria.} The complete closed-loop redesign cycle described above follows the sequence shown in Fig.~\ref{fig:overview_framework}. The QITE-related portion of Step II is further detailed in Fig.~\ref{fig:qite_workflow}.

\begin{remark}
The proposed framework introduces several approximation layers. The continuous controller and Lyapunov parameters are first restricted to a finite grid through binary encoding, so the attainable resolution depends on the calibrated interval widths and the number of assigned bits. In addition, violations of the Lyapunov conditions are penalized at finitely sampled states, and the resulting nonlinear encoded objective is approximated by a quadratic surrogate. QITE then provides an approximate low-energy solution of this surrogate. To mitigate these approximations, decoded candidates are re-evaluated using the original nonlinear cost and Lyapunov penalties before selecting the final update.
\end{remark}

\subsection{Dynamic Algorithm and Practical Implementation}
\label{subsec:algorithm_implementation}

The proposed two-step framework jointly updates controller and Lyapunov candidate parameters at successive decision epochs and applies them over the next interval, forming an online redesign loop rather than a fixed offline design.

Redesign may be periodic or conditional. Periodic updates occur every \(\Delta t\) and suit continuously or predictably changing conditions, including time-varying references, slowly varying loads, and scheduled operating points. Conditional updates are triggered by events such as disturbances, saturation, parameter mismatch, excessive tracking error, degraded transients, or violation of a prescribed Lyapunov decrease margin.

The framework also supports parallel and supervisory implementations. In parallel operation, the plant retains its current parameters while optimization runs on a separate computational unit, and updates are applied once available. In supervisory operation, the optimization layer monitors performance and updates the controller when improvement or corrective action is needed. Both forms accommodate continued plant operation during computation.

The method may be used online or offline. Online deployment requires computation fast enough for the intended redesign rate, unless the plant can be temporarily paused during updates. Offline deployment generates scheduled or scenario-dependent parameter updates in advance for subsequent operation.

Current quantum hardware limits the size of the encoded problem. Quantum simulators and classical exact or near-exact solvers therefore provide practical options for studying moderate problem dimensions. Simulator computation grows rapidly with the number of qubits, making simulation better suited to offline validation or parallel and supervisory studies than strict real-time deployment. The framework is motivated by dynamic co-design for improved closed-loop performance, while its encoded structure provides a pathway toward future implementation on fault-tolerant quantum computers.

The encoded optimization also admits alternative solvers. Quantum annealing can solve the fitted objective in QUBO or Ising form while retaining the same calibration, encoding, decoding, and nonlinear candidate evaluation steps. Compared with current gate based hardware, available annealers offer larger physical qubit counts, providing potential capacity for larger encoded problems, although the usable logical size depends on connectivity and embedding overhead. This makes annealing a complementary option for near term applications, while QITE provides the circuit based implementation investigated here.

The encoding requires
\begin{equation}
n_q=\sum_{j=1}^{n_p} n_{k,j},
\end{equation}
where \(n_p\) is the number of controller and Lyapunov candidate parameters and \(n_{k,j}\) is the number of bits assigned to parameter \(j\) at epoch \(t_k\). With a uniform resolution of \(n_b\) bits, \(n_q=n_p n_b\). The logical register size therefore grows linearly with the parameter count and bits per parameter, while the number of encoded candidates grows as \(2^{n_q}\).

For quantum annealing, embedding the logical interactions into the hardware connectivity may require additional physical qubits.

\section{Simulation Results}
\label{sec:simulation_results}

This section evaluates the proposed two-step dynamic co-design framework on three nonlinear control problems. In each case, the same online redesign procedure is used, with only the plant, controller, Lyapunov candidate, and performance index changed. At every decision epoch, the measured state is used to update the controller and {Lyapunov candidate parameters} through the two-step procedure of Section~\ref{sec:proposed_work}. The implementation is {available at \href{https://github.com/LSU-RAISE-LAB/DQCLS-NS}{GitHub} in \cite{GitHub_raiselab_code}}. {Simulations use Qiskit \(0.46.3\), \texttt{qiskit algorithms} \(0.3.1\), \texttt{VarQITE}, and noiseless statevector simulation.}

\subsection{Simulation Setup and Baselines}
\label{subsec:simulation_setup}

The numerical studies evaluate the same online co-design procedure on three nonlinear control problems while changing only the plant model, controller parameterization, Lyapunov candidate, and performance index. The main simulation settings are summarized in Table~\ref{tab:simulation_setup}. In all examples, the final parameter update is not selected directly from the fitted surrogate. Instead, the most probable encoded candidates are decoded, re-evaluated using the original nonlinear closed-loop cost with Lyapunov penalties, and compared before applying the update.

{The two consensus examples use an adaptive binary representation, assigning \(2\)--\(4\) bits to each of the five design parameters according to the calibrated interval width. Thus, the quantum register size varies between \(10\) and \(20\) qubits across redesign epochs. Wider intervals receive more bits to preserve resolution, while narrower intervals use fewer bits.} The induction motor example instead uses a fixed \(3\)-bit representation to demonstrate the framework on a more complex nonlinear system with time-varying references and uncertainty.

The main baseline in each case is a fixed parameter offline design. In this baseline, the controller and {Lyapunov candidate parameters} are selected once and then kept constant throughout the simulation. 
This comparison evaluates the overall joint tuning framework against fixed tuning, without isolating the benefits of QITE or Lyapunov candidate updates.

The selected Black-Hole settings were chosen to provide a moderate exploration--computation tradeoff. {For the two consensus examples, \(N_{\mathrm{BH}}=20\), the maximum number of iterations is \(200\), and the freezing threshold is \(\delta_j=5\). For the induction motor example, \(N_{\mathrm{BH}}=5\), the maximum number of iterations is \(10\), and \(\delta_j=10\).} These settings balance search-space contraction with the number of nonlinear closed-loop objective evaluations required at each epoch.

\begin{table*}[!t]
    \centering
    \scriptsize
    \caption{Simulation setup for the numerical studies}
    \label{tab:simulation_setup}
    \renewcommand{\arraystretch}{1.45}
    \setlength{\tabcolsep}{3.5pt}
    \resizebox{\textwidth}{!}{%
    \begin{tabular}{c|p{4.1cm}|p{4.3cm}|p{4.8cm}}
        \hline
        \textbf{Setting}
        & \textbf{First-order Consensus}
        & \textbf{Second-order Consensus}
        & \textbf{Induction Motor Drive} \\
        \hline

        \multicolumn{4}{c}{\textbf{System Description}} \\
        \hline

        \textbf{Model}
        & Five-agent nonlinear consensus with ring topology adopted from \cite{hasanzadeh2024distributed}
        & Five-agent second-order consensus with ring topology adopted from \cite{olfati2004consensus}
        & Induction motor model in the stationary reference frame adopted from \cite{farasat2014efficiency} \\
        \hline

        \textbf{Plant / operating details}
        & Nonlinear first-order agent dynamics
        & Linear and quadratic drag with coefficients \(0.5\) and \(0.05\)
        & Nominal controller model: \(R_s=2.3\), \(R_r=2.5\), \(L_m=0.24\), and \(J=0.003\); plant mismatch represented by a \(50\%\) reduction in \(L_m\) \\
        \hline

        \multicolumn{4}{c}{\textbf{Online Co-Design Variables and Search Ranges}} \\
        \hline

        \textbf{Design variables}
        & \(\alpha,\beta,k,\theta_2,\theta_4\)
        & \(K_p,K_d,\theta_{x2},\theta_{v2},\theta_{x4}\)
        & \(k_{\psi},k_{\omega},\theta_{\psi},\theta_{\omega}\) \\
        \hline

        \textbf{Initial intervals}
        & \([0,50]\), \([0,2]\), \([0,50]\), \([0,25]\), \([0,25]\)
        & \([0,50]\), \([0,50]\), \([0,50]\), \([0,40]\), \([0,20]\)
        & \([-10^2,10^3]\), \([-10^2,10^3]\), \([0.01,10^2]\), \([0.01,10^2]\) \\
        \hline

        \multicolumn{4}{c}{\textbf{Redesign and Sample-Based Cost Evaluation}} \\
        \hline

        \textbf{Redesign interval / maximum time}
        & \(\Delta t=0.25\) s; maximum time \(10\) s
        & \(\Delta t=0.5\) s; maximum time \(20\) s
        & \(\Delta t=0.2\) s; total time \(2.2\) s \\
        \hline

        {\textbf{Trajectory sampling}}
        & {Candidate-dependent closed-loop trajectory, uniformly sampled from the state measured at each redesign epoch}
        & {Candidate-dependent closed-loop trajectory, uniformly sampled from the position and velocity measured at each redesign epoch}
        & {Candidate-dependent closed-loop trajectory, uniformly sampled in absolute time to account for the time-varying references and load} \\
        \hline

        {\textbf{Evaluation horizon / samples}}
        & {\(0.25\) s with \(N_s=150\) points}
        & {\(0.25\) s with \(N_s=150\) points}
        & {\(0.05\) s with \(N_s=20\) points} \\
        \hline

        {\textbf{Sample spacing}}
        & {\(\Delta\tau_s=0.25/(150-1)=1.678\times10^{-3}\) s}
        & {\(\Delta\tau_s=0.25/(150-1)=1.678\times10^{-3}\) s}
        & {\(\Delta\tau_s=0.05/(20-1)=2.632\times10^{-3}\) s} \\
        \hline

        \textbf{Performance weights}
        & \(W_Z=1\), \(W_U=0.1\)
        & \(W_Z=1\), \(W_U=0.1\)
        & \(W_{e_{\psi}}=2\), \(W_{e_{\omega}}=10\), \(W_{e_i}=10\), \(W_U=0.01\) \\
        \hline

        {\textbf{Lyapunov-decay penalty}}
        & \multicolumn{3}{c}{
        {\(\rho_{\dot V}\int[\max(0,\dot V)]^2dt\), with \(\rho_{\dot V}=1\) and zero explicit decay margin}
        } \\
        \hline

        {\textbf{Positivity enforcement}}
        & {Structural through the disagreement-based Lyapunov form and nonnegative bounds on \(\theta_2,\theta_4\); no separate \(\rho_V\Pi_V\)}
        & {Structural through the disagreement-based Lyapunov form and nonnegative coefficient bounds; no separate \(\rho_V\Pi_V\)}
        & {Structural through the selected Lyapunov form and strictly positive lower bounds on \(\theta_{\psi},\theta_{\omega}\); no separate \(\rho_V\Pi_V\)} \\
        \hline

        {\textbf{Additional constraint penalty}}
        & \multicolumn{3}{c}{
        {No separate application-dependent penalty \(\rho_c\Pi_c\) is used in the reported numerical examples}
        } \\
        \hline

        \multicolumn{4}{c}{\textbf{Calibration, Encoding, and Encoded Optimization}} \\
        \hline

        \textbf{BH calibration}
        & \(N_{\mathrm{BH}}=20\), maximum \(200\) iterations, \(\delta_j=5\)
        & \(N_{\mathrm{BH}}=20\), maximum \(200\) iterations, \(\delta_j=5\)
        & \(N_{\mathrm{BH}}=5\), maximum \(10\) iterations, \(\delta_j=10\) \\
        \hline

        \textbf{Binary encoding}
        & Adaptive \(2\)--\(4\) bits per parameter
        & Adaptive \(2\)--\(4\) bits per parameter
        & Fixed \(3\) bits per parameter \\
        \hline

        \textbf{Number of qubits}
        & \(10\)--\(20\) qubits
        & \(10\)--\(20\) qubits
        & \(12\) qubits  \\
        \hline

        \textbf{QUBO fitting / screening}
        & Training-sample factor \(4\), minimum \(64\) bitstrings, and top-\(32\) candidate screening
        & Training-sample factor \(4\), minimum \(64\) bitstrings, and top-\(32\) candidate screening
        & Training-sample factor \(1\), minimum \(12\) bitstrings, and top-\(8\) candidate screening \\
        \hline

        \textbf{QITE settings}
        & \texttt{EfficientSU2} ansatz, \(2\) repetitions, \(\tau=3\), and \(60\) imaginary-time steps
        & \texttt{EfficientSU2} ansatz, \(2\) repetitions, \(\tau=3\), and \(60\) imaginary-time steps
        & \texttt{EfficientSU2} ansatz, \(1\) repetition, \(\tau=2\), and \(5\) imaginary-time steps \\
        \hline

        \multicolumn{4}{c}{\textbf{Stopping, Validation, and Numerical Integration}} \\
        \hline

        \textbf{Stopping / operating condition}
        & Terminate if \(\|Lx\|_2\leq10^{-8}\)
        & Terminate if \(\sqrt{\|Lx\|_2^2+\|Lv\|_2^2}\leq10^{-4}\)
        & Rotor flux reference \(0.9\), piecewise-linear speed reference, and step load torque at \(t=0.5\) s \\
        \hline

        \textbf{Numerical integration}
        & \multicolumn{3}{c}{
        \texttt{solve\_ivp} with \texttt{RK45}, relative tolerance \(10^{-6}\), and absolute tolerance \(10^{-8}\)
        } \\
        \hline

    \end{tabular}%
    }
\end{table*}

\subsection{{First-Order Nonlinear Consensus}}
\label{subsec:first_order_consensus}

For the first-order case, the closed-loop dynamics are
\begin{equation}
\dot{x}=(1-\alpha)x+(1-\beta)x^{3}-kLx,
\label{eq:first_order_consensus_closed_loop}
\end{equation}
where \(x\in\mathbb{R}^{5}\) is the stacked agent-state vector and \(L\) is the graph Laplacian. The initial condition is selected as
\begin{equation}
x(0)=
\begin{bmatrix}
2.0 & -2.5 & 3.8 & -3.2 & 0.3
\end{bmatrix}^{\top}.
\label{eq:first_order_initial_condition}
\end{equation}

{To characterize convergence to the consensus manifold, define the network average and disagreement vector as}
\begin{equation}
{
\bar{x}=\frac{1}{5}\mathbf{1}^{\top}x,
\qquad
e=x-\bar{x}\mathbf{1}.
}
\label{eq:first_order_disagreement}
\end{equation}
{The Lyapunov candidate is then selected in the disagreement coordinates as}
\begin{equation}
{
V(e)=\frac{\theta_{2}}{2}\|e\|^{2}
+\frac{\theta_{4}}{4}\sum_{i=1}^{5}e_i^{4}.
}
\label{eq:first_order_consensus_lyapunov}
\end{equation}
{This candidate vanishes for any consensus state. For the controller considered in \eqref{eq:first_order_consensus_closed_loop}, the local state-feedback terms additionally drive the consensus value toward origin.}

\begin{figure}[!t]
    \centering
    \captionsetup{font={footnotesize}}
    
    \begin{subfigure}[b]{0.49\linewidth}
        \centering
        \includegraphics[width=\linewidth]{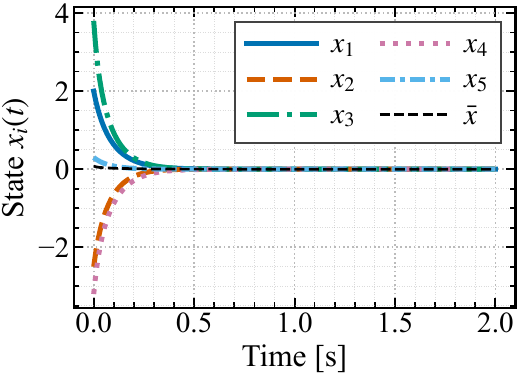}
        \caption{Agent-state trajectories}
        \label{fig:first_order_states_sub}
    \end{subfigure}
    \hfill
    \begin{subfigure}[b]{0.49\linewidth}
        \centering
        \includegraphics[width=\linewidth]{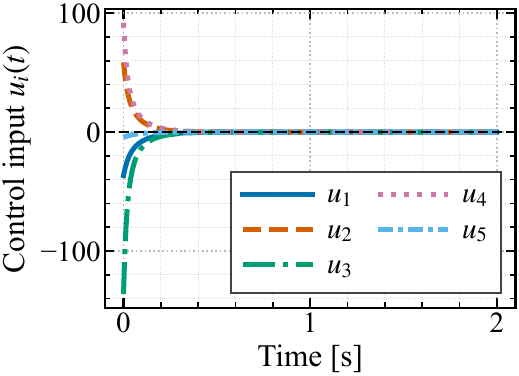}
        \caption{Control signals}
        \label{fig:first_order_control_sub}
    \end{subfigure}
    
    \vspace{0.2em}
    
    \begin{subfigure}[b]{0.49\linewidth}
        \centering
        \includegraphics[width=\linewidth]{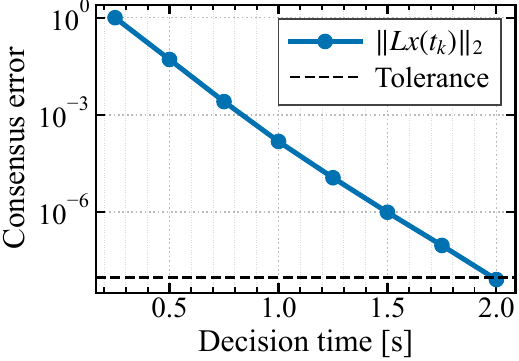}
        \caption{Consensus error history}
        \label{fig:first_order_consensus_error_sub}
    \end{subfigure}
    \hfill
    \begin{subfigure}[b]{0.49\linewidth}
        \centering
        \includegraphics[width=\linewidth]{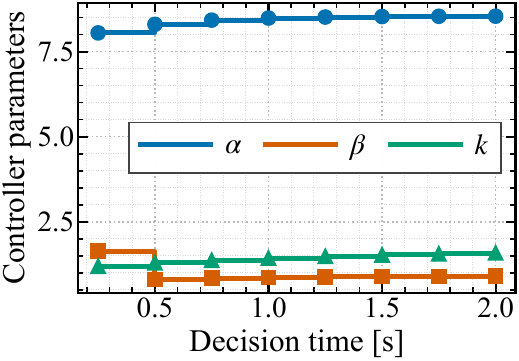}
        \caption{Controller-parameter redesign}
        \label{fig:first_order_controller_params_sub}
    \end{subfigure}
    
    \vspace{0.2em}
    
    \begin{subfigure}[b]{0.49\linewidth}
        \centering
        \includegraphics[width=\linewidth]{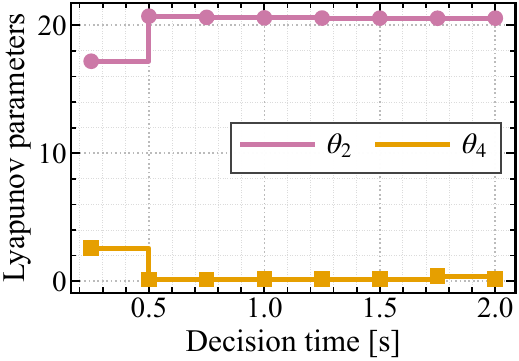}
        \caption{Lyapunov-parameter redesign}
        \label{fig:first_order_lyapunov_params_sub}
    \end{subfigure}
    \hfill
    \begin{subfigure}[b]{0.49\linewidth}
        \centering
        \includegraphics[width=\linewidth]{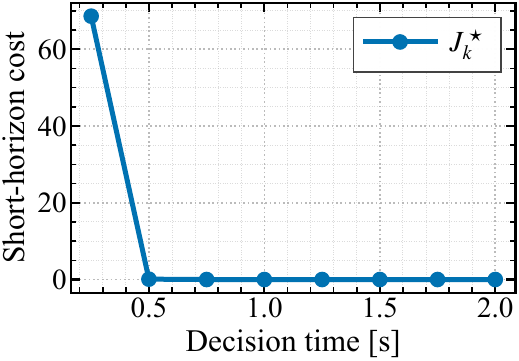}
        \caption{Short-horizon cost history}
        \label{fig:first_order_cost_history_sub}
    \end{subfigure}
    
    \caption{First-order consensus responses and online parameter redesign}
    \label{fig:first_order_consensus_all}
\end{figure}

The online co-design variables for this case are those reported in Table~\ref{tab:simulation_setup}, namely \(\alpha\), \(\beta\), \(k\), \(\theta_2\), and \(\theta_4\). Fig.~\ref{fig:first_order_consensus_all} summarizes the resulting closed-loop and redesign behavior. Fig.~\ref{fig:first_order_states_sub} shows the agent-state trajectories, where the states evolve from distinct initial conditions toward agreement. Fig.~\ref{fig:first_order_control_sub} shows the corresponding control signals, whose magnitude is larger during the initial transient and decreases as the disagreement is reduced. Fig.~\ref{fig:first_order_consensus_error_sub} reports the consensus error at the redesign epochs and confirms its monotonic reduction toward zero. {The disagreement norm decreases from \(9.924\times10^{-1}\) at the first redesign epoch to \(8.452\times10^{-9}\) at \(t=2\)~s, satisfying the prescribed consensus tolerance.} Fig.~\ref{fig:first_order_controller_params_sub} and Fig.~\ref{fig:first_order_lyapunov_params_sub} show the online evolution of the controller and Lyapunov parameters, respectively, illustrating how the proposed framework dynamically updates both sets of design variables over time. Finally, Fig.~\ref{fig:first_order_cost_history_sub} shows the short-horizon co-design cost at successive decision epochs.

{For additional validation, the complete redesign was repeated for \(\rho_{\dot V}\in\{0,0.1,1,10\}\) and \(N_s\in\{30,75,150,300\}\). All cases achieved final consensus error below \(10^{-8}\), and no positive \(\dot V\) values were observed. A separate \(1000\)-point evaluation over each prediction horizon also detected no sampled decay-condition violations.}

{For the first-order consensus case, the surrogate achieved a mean validation range-normalized RMSE of \(8.23\times10^{-5}\), validation \(R^2\approx1\), mean Spearman correlation of \(0.9971\), and mean pairwise ranking accuracy of \(0.9977\). Although exact nonlinear re-evaluation changed the surrogate rank-\(1\) candidate in two of four epochs, the retained set yielded zero screening regret and the same final consensus error, \(1.226\times10^{-10}\), as the oracle pool.}

\subsection{{Second-Order Nonlinear Consensus}}
\label{subsec:second_order_consensus}

\begin{figure}[!t]
    \centering
    \captionsetup{font={footnotesize}}
    
    \begin{subfigure}[b]{0.49\linewidth}
        \centering
        \includegraphics[width=\linewidth]{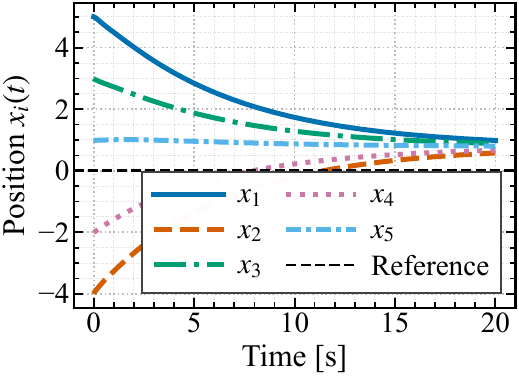}
        \caption{Position trajectories}
        \label{fig:second_order_positions_sub}
    \end{subfigure}
    \hfill
    \begin{subfigure}[b]{0.49\linewidth}
        \centering
        \includegraphics[width=\linewidth]{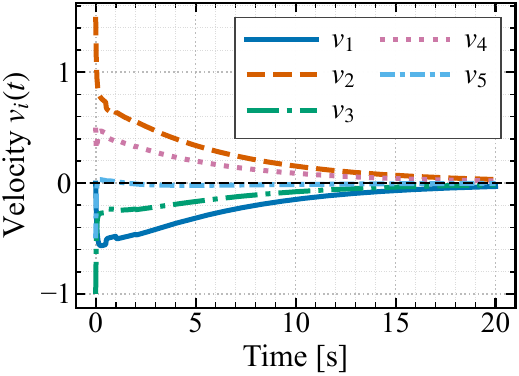}
        \caption{Velocity trajectories}
        \label{fig:second_order_velocities_sub}
    \end{subfigure}
    
    \vspace{0.2em}
    
    \begin{subfigure}[b]{0.49\linewidth}
        \centering
        \includegraphics[width=\linewidth]{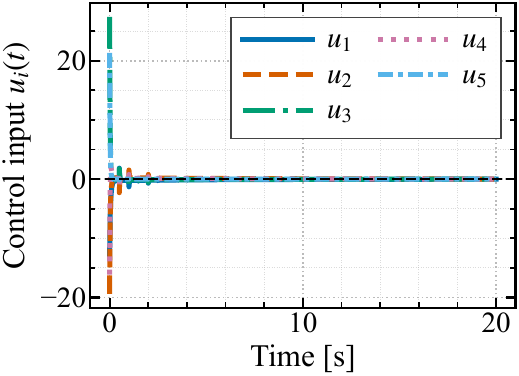}
        \caption{Control signals}
        \label{fig:second_order_control_sub}
    \end{subfigure}
    \hfill
    \begin{subfigure}[b]{0.49\linewidth}
        \centering
        \includegraphics[width=\linewidth]{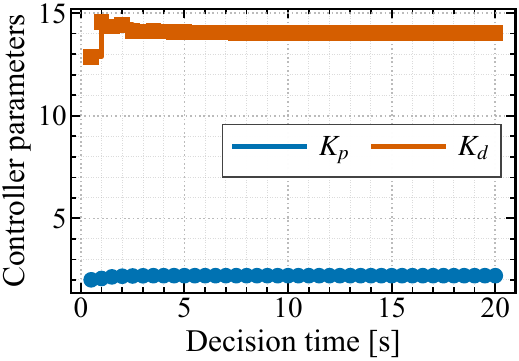}
        \caption{Controller-parameter redesign}
        \label{fig:second_order_controller_params_sub}
    \end{subfigure}
    
    \vspace{0.2em}
    
    \begin{subfigure}[b]{0.49\linewidth}
        \centering
        \includegraphics[width=\linewidth]{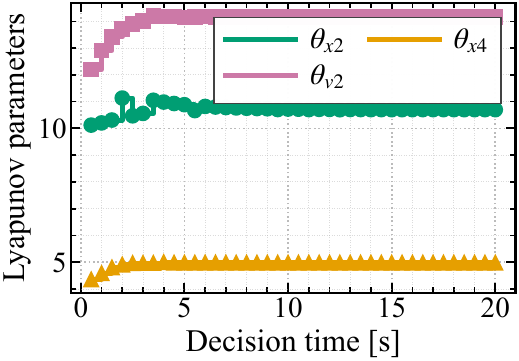}
        \caption{Lyapunov-parameter redesign}
        \label{fig:second_order_lyapunov_params_sub}
    \end{subfigure}
    \hfill
    \begin{subfigure}[b]{0.49\linewidth}
        \centering
        \includegraphics[width=\linewidth]{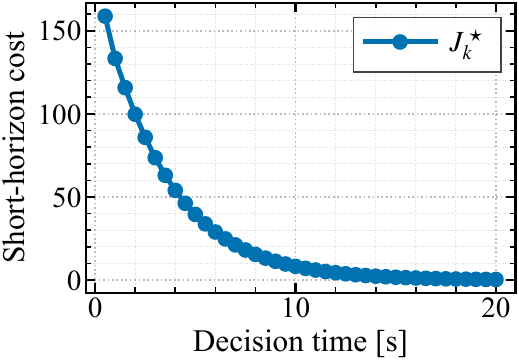}
        \caption{Short-horizon cost history}
        \label{fig:second_order_cost_history_sub}
    \end{subfigure}
    
    \caption{Second-order consensus responses and online parameter redesign}
    \label{fig:second_order_consensus_all}
\end{figure}

For the second-order case, let \(z=[x^\top\;v^\top]^\top\), where \(x\in\mathbb{R}^{5}\) and \(v\in\mathbb{R}^{5}\) denote the position and velocity vectors, respectively. The closed-loop dynamics are
\begin{equation}
\dot{x}=v,
\label{eq:second_order_consensus_xdot}
\end{equation}
\begin{equation}
\dot{v}=-a v-b|v|v-K_pLx-K_dLv,
\label{eq:second_order_consensus_vdot}
\end{equation}
where \(a\) and \(b\) are the linear and quadratic drag coefficients. The initial position and velocity vectors are initialized as
\begin{equation}
x(0)=
\begin{bmatrix}
5 & -4 & 3 & -2 & 1
\end{bmatrix}^{\top},
\label{eq:second_order_initial_position}
\end{equation}
\begin{equation}
v(0)=
\begin{bmatrix}
0 & 1.5 & -1 & 0.5 & -0.5
\end{bmatrix}^{\top}.
\label{eq:second_order_initial_velocity}
\end{equation}
Lyapunov candidate is defined in disagreement coordinates as
\begin{equation}
V(z)=\frac{\theta_{x2}}{2}\|Lx\|^{2}
+\frac{\theta_{v2}}{2}\|Lv\|^{2}
+\frac{\theta_{x4}}{4}\sum_{i=1}^{5}(Lx)_i^{4}.
\label{eq:second_order_consensus_lyapunov}
\end{equation}
The online co-design variables for this case are those reported in Table~\ref{tab:simulation_setup}, namely \(K_p\), \(K_d\), \(\theta_{x2}\), \(\theta_{v2}\), and \(\theta_{x4}\). Fig.~\ref{fig:second_order_consensus_all} summarizes the resulting closed-loop and redesign behavior. Fig.~\ref{fig:second_order_positions_sub} and Fig.~\ref{fig:second_order_velocities_sub} show the position and velocity trajectories, respectively, where both channels evolve from distinct initial conditions toward agreement under the proposed dynamic redesign mechanism. Fig.~\ref{fig:second_order_control_sub} shows the corresponding control signals. Fig.~\ref{fig:second_order_controller_params_sub} and Fig.~\ref{fig:second_order_lyapunov_params_sub} show the online evolution of the controller and Lyapunov parameters, respectively, illustrating how the proposed framework dynamically updates both sets of design variables over time. Finally, Fig.~\ref{fig:second_order_cost_history_sub} shows the short-horizon co-design cost at successive decision epochs.

\subsection{Induction Motor Drive}
\label{subsec:induction_motor_drive}

\begin{figure}[!t]
    \centering
    \captionsetup{font={footnotesize}}
    \includegraphics[width=\linewidth]{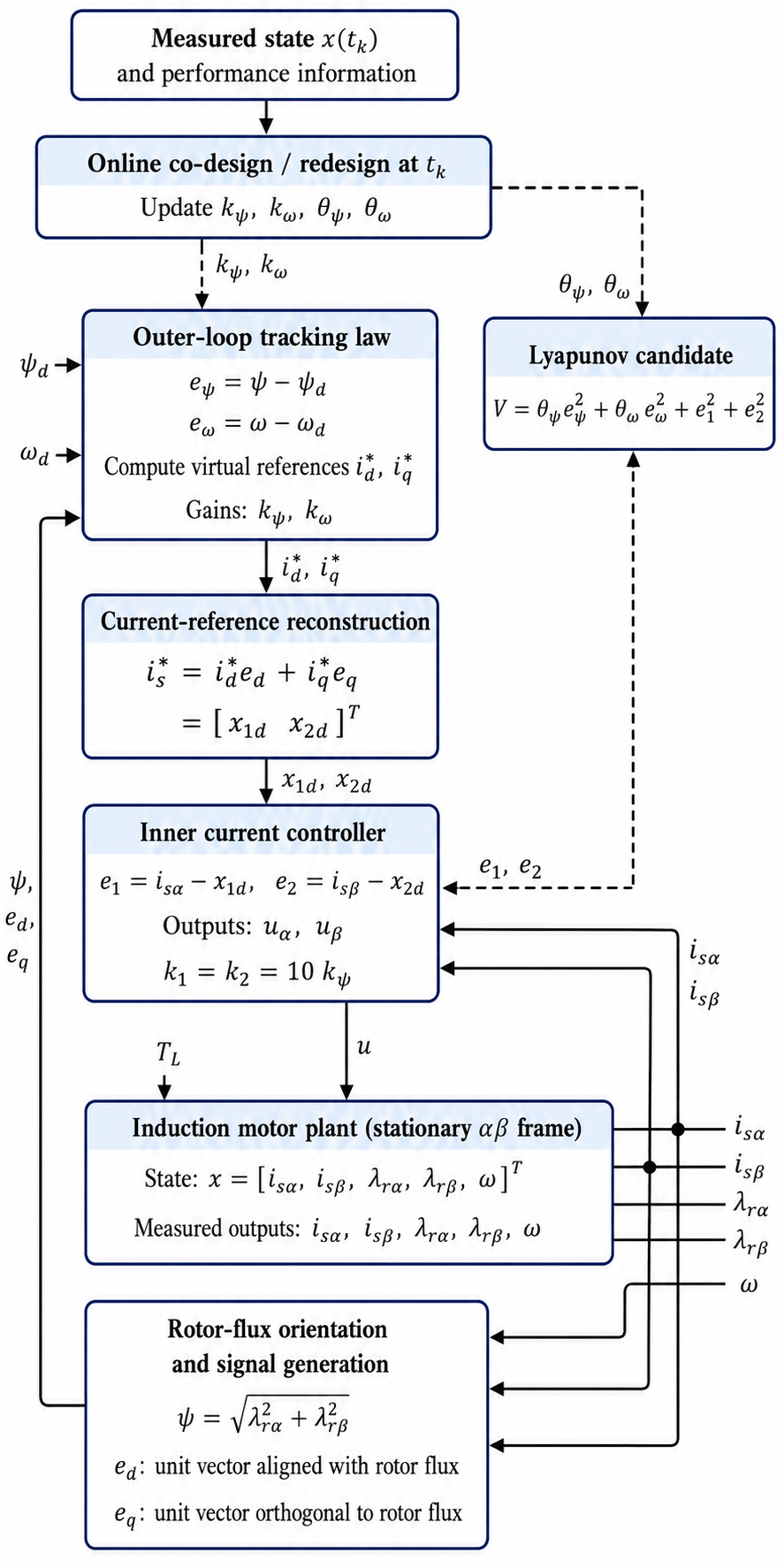}
    \caption{Induction motor drive closed-loop}
    \label{fig:im_control_architecture}
\end{figure}

The overall control architecture is shown in Fig.~\ref{fig:im_control_architecture}. The induction motor is described in the stationary reference frame, while the control law is constructed in a rotor flux-oriented frame (see \cite{farasat2014efficiency}). Induction motor state vector is defined as
\begin{equation}
x=
\begin{bmatrix}
i_{s\alpha} & i_{s\beta} & \lambda_{r\alpha} & \lambda_{r\beta} & \omega
\end{bmatrix}^{\top},
\end{equation}
where \(i_{s\alpha}\) and \(i_{s\beta}\) are the \(\alpha\)-\(\beta\) stator current components, \(\lambda_{r\alpha}\) and \(\lambda_{r\beta}\) are the \(\alpha\)-\(\beta\) rotor flux components, and \(\omega\) is the motor speed. The motor parameters are the stator resistance \(R_s\), rotor resistance \(R_r\), stator inductance \(L_s\), rotor inductance \(L_r\), magnetizing inductance \(L_m\), rotor inertia \(J\), pole pair number \(p_p\), and load torque \(T_L\). The subscript \(p\) denotes plant side quantities, which are used in the simulated motor, whereas the subscript \(c\) denotes the nominal controller side model. Two parameter sets are considered to represent common electric drive parameter variations: the stator resistance may increase as the winding temperature rises, while the magnetizing inductance may decrease when the motor core approaches saturation. Therefore, the plant side parameters capture the mismatched motor behavior, whereas the controller parameters represent nominal model used in the control law.

The compact coefficients used in the stationary frame model are constructed from the physical parameters as follows:
\[
L_{\sigma,p}=L_{s,p}-\frac{L_{m,p}^{2}}{L_{r,p}},
\qquad
a_{s,p}=\frac{R_{s,p}}{L_{\sigma,p}}+\frac{R_{r,p}L_{m,p}^{2}}{L_{\sigma,p}L_{r,p}^{2}},
\]
\[
b_{s,p}=\frac{R_{r,p}L_{m,p}}{L_{\sigma,p}L_{r,p}^{2}},
\qquad
c_{s,p}=\frac{L_{m,p}}{L_{\sigma,p}L_{r,p}},
\]
\[
\alpha_p=\frac{R_{r,p}L_{m,p}}{L_{r,p}},
\qquad
\beta_p=-\frac{R_{r,p}}{L_{r,p}},
\qquad
\gamma_p=\frac{3p_pL_{m,p}}{2JL_{r,p}}.
\]
The same definitions are used for the controller side coefficients \(L_{\sigma,c}\), \(a_{s,c}\), \(b_{s,c}\), \(c_{s,c}\), \(\alpha_c\), \(\beta_c\), and \(\gamma_c\) by replacing the plant parameters with their nominal controller model values.

Using these coefficients, the induction motor dynamics in the stationary reference frame are
\begin{equation}
\dot{i}_{s\alpha}
=
-a_{s,p} i_{s\alpha}
+b_{s,p} \lambda_{r\alpha}
+c_{s,p}\omega \lambda_{r\beta}
+\frac{1}{L_{\sigma,p}}u_{\alpha},
\label{eq:im_closed_loop_x1_case2}
\end{equation}
\begin{equation}
\dot{i}_{s\beta}
=
-a_{s,p} i_{s\beta}
+b_{s,p} \lambda_{r\beta}
-c_{s,p}\omega \lambda_{r\alpha}
+\frac{1}{L_{\sigma,p}}u_{\beta},
\label{eq:im_closed_loop_x2_case2}
\end{equation}
\begin{equation}
\dot{\lambda}_{r\alpha}
=
\alpha_p i_{s\alpha}
+\beta_p \lambda_{r\alpha}
-\omega \lambda_{r\beta},
\label{eq:im_closed_loop_x3_case2}
\end{equation}
\begin{equation}
\dot{\lambda}_{r\beta}
=
\alpha_p i_{s\beta}
+\beta_p \lambda_{r\beta}
+\omega \lambda_{r\alpha},
\label{eq:im_closed_loop_x4_case2}
\end{equation}
\begin{equation}
\dot{\omega}
=
\gamma_p \psi i_q
-\frac{1}{J}T_L,
\label{eq:im_closed_loop_x5_case2}
\end{equation}
where
\begin{equation}
\psi=\sqrt{\lambda_{r\alpha}^{2}+\lambda_{r\beta}^{2}}
\label{eq:im_flux_magnitude_case2}
\end{equation}
is the rotor flux magnitude.

The motor is initialized from
\begin{equation}
x(0)=
\begin{bmatrix}
0 & 0 & 0.9 & 0 & 0
\end{bmatrix}^{\top},
\label{eq:im_initial_condition}
\end{equation}
so that the rotor flux magnitude starts at its nominal reference value and the motor speed starts from rest.

The controller is constructed in a rotor flux-oriented frame. Let
\begin{equation}
e_d=\frac{1}{\psi}
\begin{bmatrix}
\lambda_{r\alpha}\\
\lambda_{r\beta}
\end{bmatrix},
\qquad
e_q=\frac{1}{\psi}
\begin{bmatrix}
-\lambda_{r\beta}\\
\lambda_{r\alpha}
\end{bmatrix},
\label{eq:im_ed_eq_case2}
\end{equation}
where \(e_d\) is the unit vector aligned with the rotor flux direction and \(e_q\) is the orthogonal unit vector. The flux and speed tracking errors are defined as
\begin{equation}
e_{\psi}=\psi-\psi_d,
\qquad
e_{\omega}=\omega-\omega_d,
\label{eq:im_outer_errors_case2}
\end{equation}
where \(\psi_d\) and \(\omega_d\) are the rotor flux and speed references, respectively.

The outer loop generates the desired \(d\)- and \(q\)-axis current components through
\begin{equation}
i_d^\star
=
\frac{\dot{\psi}_d-\beta_c \psi-k_{\psi}e_{\psi}}{\alpha_c},
\label{eq:im_id_star_case2}
\end{equation}
\begin{equation}
i_q^\star
=
\frac{\dot{\omega}_d+\frac{1}{J}T_L-k_{\omega}e_{\omega}}{\gamma_c \psi_{\mathrm{eff}}},
\qquad
\psi_{\mathrm{eff}}=\max(\psi,\psi_{\mathrm{floor}}),
\label{eq:im_iq_star_case2}
\end{equation}
where \(k_{\psi}\) and \(k_{\omega}\) are the outer loop gains and \(\psi_{\mathrm{floor}}>0\) is a small lower bound introduced to avoid division by very small flux magnitudes.

The desired stator current vector in the stationary frame is then obtained as
\begin{equation}
i_s^\star=i_d^\star e_d+i_q^\star e_q.
\label{eq:im_is_star_case2}
\end{equation}
Equivalently, if
\begin{equation}
i_s^\star=
\begin{bmatrix}
x_{1d} & x_{2d}
\end{bmatrix}^{\top},
\end{equation}
then the inner-loop current-tracking errors are
\begin{equation}
e_1=i_{s\alpha}-x_{1d},
\qquad
e_2=i_{s\beta}-x_{2d}.
\label{eq:im_inner_errors_case2}
\end{equation}

The control inputs applied to the plant are the \(\alpha\)-\(\beta\) stator voltage components \(u_{\alpha}\) and \(u_{\beta}\), selected as
\begin{equation}
u_{\alpha}
=
L_{\sigma,c}
\Big(
\dot{x}_{1d}
+a_{s,c}i_{s\alpha}
-b_{s,c}\lambda_{r\alpha}
-c_{s,c}\omega\lambda_{r\beta}
-k_1 e_1
\Big),
\label{eq:im_u_alpha_case2}
\end{equation}
\begin{equation}
u_{\beta}
=
L_{\sigma,c}
\Big(
\dot{x}_{2d}
+a_{s,c}i_{s\beta}
-b_{s,c}\lambda_{r\beta}
+c_{s,c}\omega\lambda_{r\alpha}
-k_2 e_2
\Big).
\label{eq:im_u_beta_case2}
\end{equation}

The inner-loop gains \(k_1\) and \(k_2\) are tied to the flux loop gain to maintain a faster current-tracking loop than the outer flux and speed loops. Specifically, we use
\begin{equation}
k_1=10k_{\psi},
\qquad
k_2=10k_{\psi}.
\label{eq:im_inner_gains_case2}
\end{equation}
This provides a practical time-scale separation, so that the stator current errors \(e_1\) and \(e_2\) decay faster than the outer loop errors while avoiding unnecessarily large voltage commands.
Thus, the outer loop determines the desired flux and torque-producing currents, whereas the inner loop drives the stator currents toward these references through the voltage inputs.

The Lyapunov candidate is chosen as
\begin{equation}
V(x)
=
\theta_{\psi} e_{\psi}^{2}
+\theta_{\omega} e_{\omega}^{2}
+e_{1}^{2}
+e_{2}^{2},
\label{eq:im_lyapunov_candidate_case2}
\end{equation}
where \(\theta_{\psi}\) and \(\theta_{\omega}\) are adjustable Lyapunov weights. Accordingly, the online co-design vector is
\begin{equation}
p_k=
\begin{bmatrix}
k_{\psi,k} & k_{\omega,k} & \theta_{\psi,k} & \theta_{\omega,k}
\end{bmatrix}^{\top}.
\end{equation}
At each epoch, the proposed framework updates these quantities using the measured motor state, as illustrated in Fig.~\ref{fig:im_control_architecture}.

\begin{figure}[!t]
    \centering
    \captionsetup{font={footnotesize}}

    \begin{subfigure}[b]{0.49\linewidth}
        \centering
        \includegraphics[width=\linewidth]{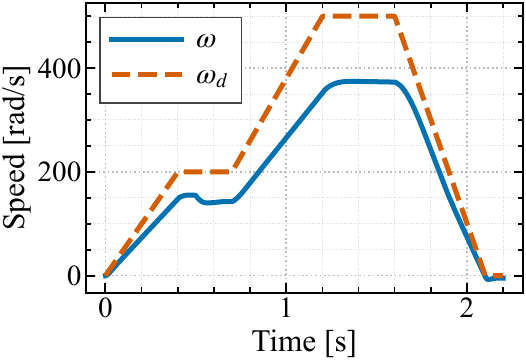}
        \caption{Motor speed response}
        \label{fig:im_fixed_gain_speed_sub}
    \end{subfigure}
    \hfill
    \begin{subfigure}[b]{0.49\linewidth}
        \centering
        \includegraphics[width=\linewidth]{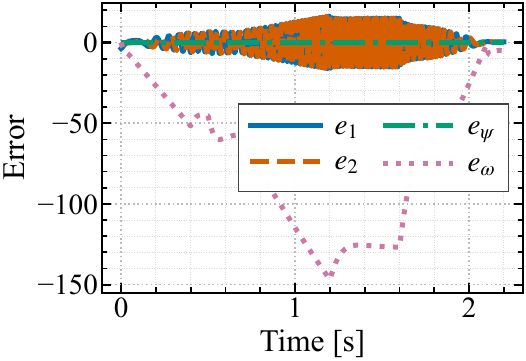}
        \caption{Tracking errors}
        \label{fig:im_fixed_gain_errors_sub}
    \end{subfigure}

    \caption{Fixed gain controller under the \(50\%\) \(L_{m,p}\)-mismatch}
    \label{fig:im_fixed_gain_baseline}
\end{figure}

Before evaluating the proposed online co-design strategy, we first consider a fixed gain implementation of the same rotor flux-oriented controller. In this baseline case, the controller gains are kept constant over the full simulation horizon, with
\begin{equation}
k_{\psi}=100,
\qquad
k_{\omega}=100,
\label{eq:im_fixed_gains_case2}
\end{equation}
and no online redesign is performed. The purpose of this baseline is to show that the underlying controller can be tuned to track well under nominal conditions, while also demonstrating that fixed nominal gains may not be sufficiently robust under severe parameter mismatch. In preliminary nominal simulations, i.e., when the controller side and plant side parameters match, these fixed gains were found to provide satisfactory speed tracking and small tracking errors. However, when the same fixed gain controller is applied to the saturation mismatch case with a \(50\%\) reduction in the plant side magnetizing inductance \(L_{m,p}\), the tracking performance deteriorates significantly. In particular, the speed response no longer follows the reference accurately, and the flux and speed tracking errors become large, as shown in Fig.~\ref{fig:im_fixed_gain_baseline}. This baseline demonstrates that a fixed gain controller tuned for nominal operation may not preserve acceptable performance under strong magnetic saturation.

\begin{figure}[!t]
    \centering
    \captionsetup{font={footnotesize}}
    
    \begin{subfigure}[b]{0.49\linewidth}
        \centering
        \includegraphics[width=\linewidth]{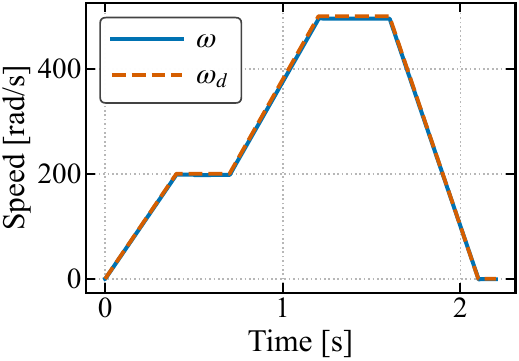}
        \caption{Motor speed response}
        \label{fig:im_speed_sub}
    \end{subfigure}
    \hfill
    \begin{subfigure}[b]{0.49\linewidth}
        \centering
        \includegraphics[width=\linewidth]{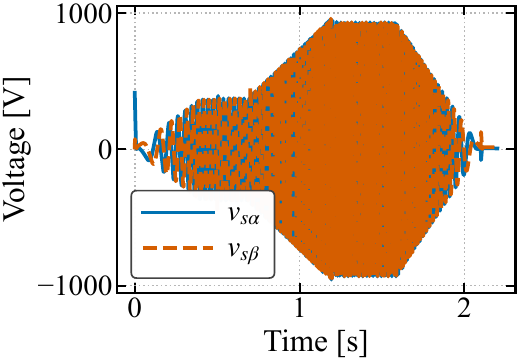}
        \caption{Input voltages}
        \label{fig:im_control_sub}
    \end{subfigure}
    
    \vspace{0.2em}
    
    \begin{subfigure}[b]{0.49\linewidth}
        \centering
        \includegraphics[width=\linewidth]{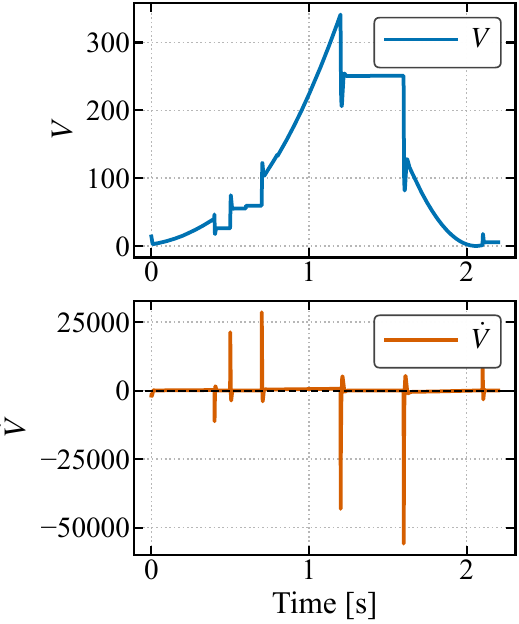}
        \caption{Lyapunov candidate }
        \label{fig:im_lyapunov_sub}
    \end{subfigure}
    \hfill
    \begin{subfigure}[b]{0.49\linewidth}
        \centering
        \includegraphics[width=\linewidth]{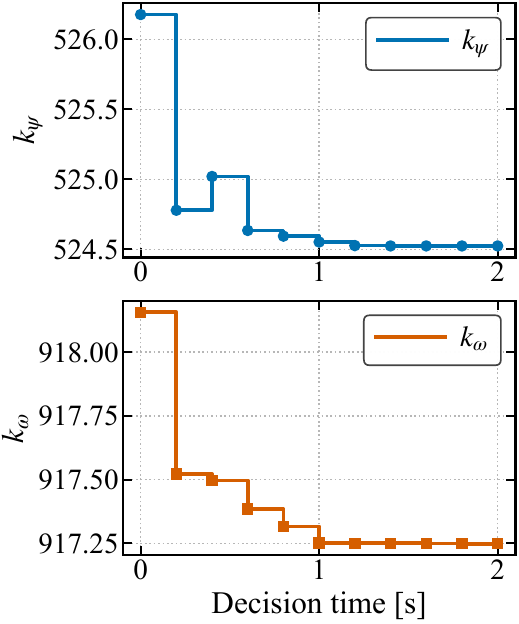}
        \caption{Controller-parameter redesign}
        \label{fig:im_controller_params_sub}
    \end{subfigure}
    
    \vspace{0.2em}
    
    \begin{subfigure}[b]{0.49\linewidth}
        \centering
        \includegraphics[width=\linewidth]{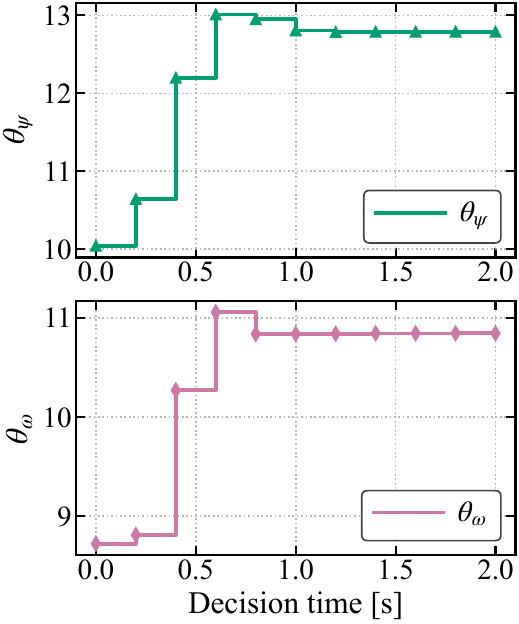}
        \caption{Lyapunov-parameter redesign}
        \label{fig:im_lyapunov_params_sub}
    \end{subfigure}
    \hfill
    \begin{subfigure}[b]{0.49\linewidth}
        \centering
        \includegraphics[width=\linewidth]{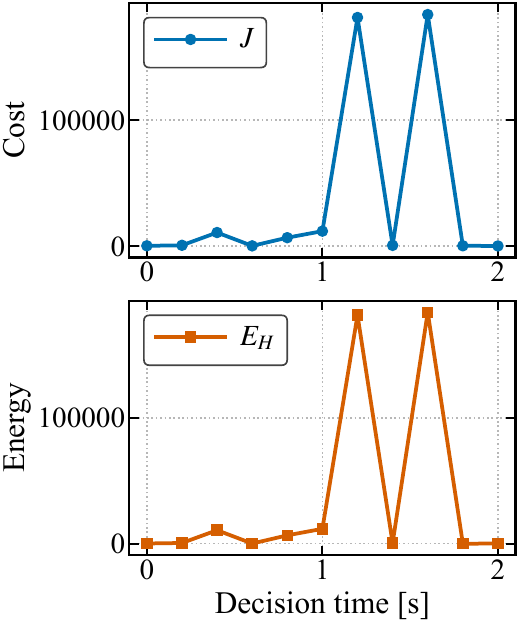}
        \caption{Short-horizon cost history}
        \label{fig:im_cost_history_sub}
    \end{subfigure}
    
    \caption{Closed-loop responses and online design-variable evolution}
    \label{fig:im_drive_all}
\end{figure}

After establishing the fixed gain baseline, we evaluate the proposed online co-design method under the same saturation mismatch scenario. The plant is operated with a \(50\%\) reduction in the plant side magnetizing inductance \(L_{m,p}\), while the controller is initialized using the nominal controller side model. Fig.~\ref{fig:im_drive_all} summarizes the closed-loop behavior obtained with the proposed online co-design method. In contrast to the fixed gain baseline in Fig.~\ref{fig:im_fixed_gain_speed_sub}, the online method preserves accurate speed tracking over the full operating horizon despite the saturation mismatch and the step load disturbance as shown in Fig.~\ref{fig:im_speed_sub}. 

Fig.~\ref{fig:im_control_sub} shows the stationary frame control voltages, whose amplitudes vary with the acceleration, cruising, and deceleration phases. Fig.~\ref{fig:im_lyapunov_sub} shows the piecewise Lyapunov response caused by the discrete parameter updates, while Figs.~\ref{fig:im_controller_params_sub} and \ref{fig:im_lyapunov_params_sub} show that the controller gains require only minor corrections, whereas the Lyapunov parameters vary more noticeably. Fig.~\ref{fig:im_cost_history_sub} shows that the short horizon cost increases during the more demanding speed transition intervals.

\subsection{Additional Sensitivity and Computational Studies}
\label{subsec:additional_studies}

To further evaluate the proposed framework beyond the main case studies, additional analyses are conducted to investigate encoded-solution quality, the effect of binary resolution, computational burden, Black-Hole calibration sensitivity, and sampled Lyapunov-condition satisfaction.

\subsubsection{{Encoded-Solution Quality}}

To assess the encoded-solution quality, the same fitted 12-variable QUBO was solved at each of the 11 redesign epochs using QITE and Gurobi as an exact binary MIQP solver. Gurobi was configured with zero optimality gap and certified global optimality at every epoch. QITE selected the same bitstring as Gurobi in 8 of the 11 epochs. In the remaining three epochs, the Hamming distances were one, one, and two, giving a mean Hamming distance of \(0.364\). Despite these alternative bitstrings, the decoded candidates produced identical re-evaluated nonlinear closed-loop costs at all epochs, with zero mean and maximum absolute cost residuals. The mean absolute Hamiltonian energy residual was \(3.43\times10^{-12}\), and the maximum was \(2.03\times10^{-11}\), indicating only numerical precision differences. Thus, for this induction motor case, QITE recovered encoded solutions equivalent in nonlinear cost to the globally optimal Gurobi solutions at every epoch.

\subsubsection{Binary-Resolution Sensitivity}

To evaluate the effect of finite binary resolution, this case was repeated using \(n_b=2,3,\ldots,8\) bits for each of the four co-design parameters. For a fair comparison, all plant parameters, controller structure, initial conditions, reference profiles, load disturbance, cost weights, Black-Hole settings, decision intervals, and random seeds were kept unchanged. At each redesign epoch, the Black-Hole stage first contracted the continuous parameter intervals. Each contracted interval was then discretized using the selected number of bits, the sampled quadratic surrogate was constructed over the resulting binary space, and the QUBO was solved using the same classical local search solver. A classical solver was used in this study to isolate the effect of binary discretization from the additional approximation associated with QITE. The decoded candidates were subsequently evaluated using the original nonlinear short horizon model, and the selected parameters were applied to the closed-loop system over the next decision interval.

For each binary resolution, the complete \(2.2\)-s closed-loop simulation was performed and the resulting state, speed, flux, current, control, and Lyapunov trajectories were recorded. The comparison included the final, mean, RMS, and maximum tracking error norms; the individual flux, speed, and current error measures; the total closed-loop cost; the maximum and integrated Lyapunov penalties; the control effort; the peak currents and voltages; and the Hamiltonian fitting and QUBO solution times. All resolutions completed successfully and produced nearly identical closed-loop behavior. The final tracking error norm remained approximately \(2.139\), the RMS tracking error norm remained near \(11.53\), and the maximum tracking error norm remained near \(19.98\). The flux error RMS remained approximately \(1.81\times10^{-2}\), while the speed error, current error, control effort, peak current, and peak voltage measures changed only marginally. The total cost and Lyapunov penalty values varied nonmonotonically because the final solution is influenced not only by the binary grid, but also by sampled surrogate fitting and approximate QUBO solution. Overall, the results show that the \(3\)-bit representation used in the main study provides sufficient resolution after interval contraction, whereas increasing the resolution beyond \(3\) bits enlarges the QUBO and increases computation time without producing an improvement in closed-loop performance.

\subsubsection{{Computational Burden and Black-Hole Sensitivity}}

A stage-wise timing study was also performed for the induction motor case using the baseline \(3\)-bit representation, corresponding to \(12\) qubits. One redesign epoch required \(48.7211\pm0.3245\)~s, of which the simulated QITE stage accounted for \(48.0854\pm0.3662\)~s. In comparison, the Black-Hole calibration and Hamiltonian fitting required \(0.1835\pm0.0218\)~s and \(0.3961\pm0.0478\)~s, respectively. The approximately \(48.7\)~s computation time per redesign cannot meet the prescribed \(0.2\)~s update interval in real time. Simulated QITE accounts for approximately \(98.7\%\) of the total runtime and takes about \(76\) times as long as all remaining stages combined. This identifies simulated QITE as the main computational bottleneck and a possible target for future improvement, without establishing a computational advantage.

The effects of binary resolution and Black-Hole iterations were also examined. Increasing the resolution from \(2\) to \(3\) bits per parameter increased the encoded problem from \(8\) to \(12\) qubits and increased the mean QITE simulation time from \(12.0704\)~s to \(49.3017\)~s. Increasing the maximum Black-Hole iterations from \(2\) to \(20\) increased its average runtime from \(0.0464\)~s to \(0.1823\)~s, while the interval contraction benefit became marginal beyond approximately \(10\) iterations.

\subsubsection{{Sampled Lyapunov-Condition Assessment}}

The maximum sampled Lyapunov constraint violations were also evaluated over the applied closed-loop trajectories. For the first-order and second-order consensus cases, both the positivity violation, \(\max_{k,i}\max(0,-V_{k,i})\), and the decay violation, \(\max_{k,i}\max(0,\dot V_{k,i})\), were zero, with maximum sampled derivatives of \(-7.46\times10^{-15}\) and \(-2.58\), respectively. For the induction motor case, the positivity violation was zero and the minimum sampled Lyapunov value was \(1.31\times10^{-1}\); however, the maximum decay violation was \(2.86\times10^{4}\),
with \(\dot V>0\) at \(1073\) of \(2190\) sampled points, meaning that sampled Lyapunov decay fails at these points. Thus, consensus cases satisfied the sampled Lyapunov conditions, whereas the motor case preserved positivity but did not achieve sampled decay, confirming that its Lyapunov term acts as a soft penalty.

\section{Conclusion}
\label{sec:conclusion}

This paper presented a two-step dynamic quantum-assisted framework for online co-design of controller parameters and {Lyapunov candidate parameters} in nonlinear closed-loop systems. The proposed method combines Black-Hole search space calibration, encoded optimization, and QITE-based quantum search to reduce the online redesign problem to a locally calibrated binary search while still selecting controller and {Lyapunov candidate parameters} through exact nonlinear closed-loop re-evaluation. A complete implementation was developed and evaluated on first-order nonlinear consensus, second-order nonlinear consensus, and induction motor drive examples using the same overall framework with only problem-dependent changes in the dynamics, controller parameterization, Lyapunov candidate, and cost function.
The framework applies across different nonlinear systems, with improvements over fixed designs attributable to the overall simulated procedure. Separate benefits of QITE and Lyapunov candidate updates are not isolated. The consensus examples satisfied positivity and decay conditions at the evaluated samples, whereas the induction motor example maintained sampled positivity but exhibited decay violations. The Lyapunov penalties guide parameter selection without providing a stability certificate over a continuous domain or across parameter updates.
 The simulation codes used in this work are made publicly available to support reproducibility.

\bibliographystyle{IEEEtran}
\bibliography{refs}

\end{document}